\documentclass[11pt]{article}

\usepackage{amsmath}
\usepackage{enumitem}
\usepackage{bm}
\usepackage{graphicx}
\usepackage{epstopdf}
\usepackage{caption}
\usepackage[labelformat=simple]{subcaption}
\usepackage[titletoc,toc]{appendix}
\numberwithin{equation}{section}

\usepackage{amssymb, graphics}

\usepackage{units} 

\def\l({\left(}
\def\r){\right)}
\def\[{\left[}
\def\]{\right]}
\def\{{\left\lbrace}
\def\}{\right\rbrace}
\def\tr{{\rm Tr}}
\def\n{\\ \noindent \newline}

\begin{document}

\renewcommand{\thefootnote}{\fnsymbol{footnote}}

\thispagestyle{empty}
\vspace{-.5in}
\begin{flushright}
OSU-HEP-16-09 \\
\end{flushright}

\vspace*{1.0cm}
\begin{center}
\LARGE {\bf \Large Radiative Electroweak Symmetry Breaking \\ in Standard Model Extensions}
\end{center}

\begin{center}

{\large \bf K.S. Babu$^a$\footnote{Email:
kaladi.babu@okstate.edu}, Ilia Gogoladze$^b$\footnote{Email: ilia@bartol.udel.edu} and S. Khan$^a$\footnote{Email:
saki.khan@okstate.edu}}\n
\end{center}

\vspace{-0.7in}
\begin{center}
\it $^a $Department of Physics, Oklahoma State University, Stillwater, Oklahoma 74078, USA \\
\it $^b $Bartol Research Institute, Department of Physics and Astronomy,
University of Delaware, Newark, DE 19716, USA
\end{center}
\vspace{.5in}
\renewcommand{\thefootnote}{\arabic{footnote}}
\setcounter{footnote}{0}

\begin{abstract}
We study the possibility of radiative electroweak symmetry breaking  where loop corrections to the mass parameter of the Higgs boson trigger the symmetry breaking in various extensions of the Standard Model (SM).  Although the mechanism fails in the SM, it is shown to be quite successful in several extensions which share a common feature of having an additional scalar around the TeV scale.  The positive Higgs mass parameter at a high energy scale is turned negative in the renormalization group flow to lower energy by the cross couplings between the scalars in the Higgs potential.  The type-II seesaw model with a TeV scale weak scalar triplet, a two-loop radiative neutrino mass model with new scalars at the TeV scale, the inert doublet model, scalar singlet dark matter model, and a universal seesaw model with an additional $U(1)$ broken at the TeV scale are studied and shown to exhibit successful radiative electroweak symmetry breaking. 
\end{abstract}
\newpage
\section{Introduction}
Discovery of the Higgs boson as predicted by the Standard Model (SM) by the ATLAS and the CMS experiments became the moment of triumph for particle physics \cite{higgsdiscovery, Aad:2015zhl}. Such a historic discovery together with decades of eletroweak precision data have well established the validity of SM up to accessable energies. However, there is no verified explanation of the origin of the small neutrino masses and no viable candidate for the dark matter in the SM. Due to these unwavering issues, various extensions of SM have been proposed. The secret of neutrino masses may lie in some form of seesaw mechanism, where a SM singlet right-handed neutrinos with large Majorana masses cause the light neutrino masses (Type-I seesaw)  \cite{seesawmodel} or a SM weak scalar triplet with a tiny induced vacuum expectation value (VEV) generates the small neutrino masses (Type-II seesaw) \cite{typeII}. If neutrino masses are generated as loop corrections, the masses will naturally be suppressed and such extensions of SM are both theoretically well motivated and phenomenologically viable \cite{Zee:1980ai,Babu:1988ki}. 

Searches for a stable dark matter candidate have also been motivation for various extensions of the SM. Some form of symmetry usually stabilizes the dark matter. Simple discrete symmetries such as R-parity in supersymmetric models \cite{Farrar:1978xj} can perform an excellent job of preventing the particle from decaying. Kaluza-Klein parity \cite{Cheng:2002iz} in universal extra dimension models and T-parity in the littlest Higgs models \cite{Tparity} can stabilize the lightest particle; turning them into promising dark matter candidates. A similar role is played by a $\mathbb{Z}_2$ symmetry for the case of inert doublet models \cite{Deshpande:1977rw} or scotogenic models \cite{Ma:2006km}. SM extended by a scalar singlet carrying a discrete $\mathbb{Z}_2$ parity is yet another example for a simple dark matter model. Instead of being an adhoc symmetry, this $\mathbb{Z}_2$ symmetry can be a remnant of the $(B-L)$ generator of $SO(10)$ grand unified theories (GUTs) \cite{z2so10}. 

$SO(10)$ GUTs provide one of the most lucrative frameworks, where one can incorporate many of the aforementioned extensions of SM along with a beautiful unified picture of SM gauge couplings. Among the classes of $SO(10)$ GUTs, supersymmetric versions have multiple features such as successful unification of gauge couplings and natural dark matter candidate owing to an automatic R-parity, while it solves the gauge hierarchy problem based on symmetry principle. In addition, supersymmetric models offer a mechanism for triggering electroweak symmetry breaking (EWSB) via radiative effects \cite{SUSYREWSB}. In this scenario, the positive mass parameter of the Higgs boson at high energy becomes negative at low energy due to the renormalization group flow which dictates how the parameters evolve with scale. 

The purpose of this paper is to explore the possibility of radiative EWSB breaking in non-supersymmetric (non-SUSY) extensions of the SM. Since this is an attractive mechanism to trigger EWSB, checking its viability in non-SUSY models is of great interest. As we argue below, such a radiative EWSB may be necessary in certain unified theories which have two stages of symmetry breaking. In a general context a positive mass parameter for the Higgs boson turning negative also enhances the available parameter space.

In some extensions of SM, the Higgs boson is a part of the larger multiplet, which breaks some higher symmetry. This occurs in trinification models based on $SU(3)_C \times SU(3)_L \times SU(3)_R$ gauge symmetry \cite{trinification01, tri02} broken by two $(1,3, 3^*)$ Higgs multiplets. These multiplets contain SM singlet components which acquire large VEVs breaking the gauge symmetry down to SM. The same $(1,3, 3^*)$ multiplets also contain the Higgs boson of the SM which should develop a negative squared mass to trigger EWSB. Consistency of the high scale symmetry breaking, however, would demand that all physical Higgs bosons, including the SM Higgs, have positive squared masses at a high energy. One could introduce new Higgs fields to break the electroweak symmetry in which case the model looses its minimality and predictivity. For such class of models one might employ radiative loop corrections to turn Higgs mass parameter negative at low energy from positive value it obtained at high energy and thus cause eletroweak symmetry breaking. A second example is provided by a class of $SO(10)$ models with the symmetry breaking sectors containing $\overline{126}_H$ along with either a $45_H$ or a $210_H$ \cite{Babu:2016cri} where flavor mixing is induced by vector-like fermions in the $16+\overline{16}$ representation. In such models, the SM singlet from $\overline{126}_H$ acquires a GUT scale VEV, breaking $SO(10)$ down to $SU(5)$. The $\overline{126}_H$ also contains a SM Higgs doublet which must have positive squared mass at the GUT scale. This positive mass term can turn negative at low energy due to renormalization group flow. Similar arguments can be applied for the case where a SM singlet of $144$-representation breaks $SO(10)$ down to SM \cite{Babu:2005gx}. The Higgs doublet is also part of $144$, which should have a positive squared mass at the GUT scale.

In this paper we explore the possibility of radiative electroweak symmetry breaking in several popular extensions of the SM. The mechanism fails in the SM, as reviewed in Sec. \ref{sec:SM and type-I}. In type-I seesaw models which includes right-handed neutrinos to the spectrum of SM, radiative EWSB is not achieved - in fact the effect of $\nu_R$ fields is to provide {\it positive} corrections to the Higgs mass parameter in evolving from high to low energies. The situation is different in type-II seesaw models which contain a weak scalar triplet, if the mass of the triplet is around the TeV scale. The key difference is the cross coupling between the SM Higgs boson and an additional scalar field in the Higgs potential. The need for this additional scalar field to be at the TeV scale arises from the needed magnitude of the $\mu_\phi^2$ parameter: $\mu_\phi^2 = -(88 \; \text{GeV})^2$. As the correction to $\mu_\phi^2$ from the scalar cross coupling grows as $-(m_\Delta^2)$ of the new scalar, this scalar should not be much heavier than a TeV, assuming that the quartic cross couplings are not extremely weak.

Radiative mass generation is a popular mechanism for neutrino masses where one assumes new scalars at the TeV scale for lepton number violation. Such models are amenable to radiative EWSB. Dark matter models employing a scalar singlet or an inert doublet also exhibit radiative EWSB. Finally, we propose and analyze a universal seesaw model wherein a new $U(1)$ symmetry is broken at TeV scale, which also shows radiative EWSB. 

In non-supersymmtric extensions of the SM such as the ones studied here, the gauge hierarchy problem has to be somehow solved. Here we simply assume that this is done by fine-tuning. The dimensionful parameters of the SM extensions are given by
\begin{equation}
\mathcal{L}_{SM}=\Lambda^4_{cos}+\Lambda^2 \mu_\phi^2 + \cdots
\end{equation}
where the $\cdots$ donate mass parameters for additional scalar fields that may be present and $\Lambda_{cos}$ is the cosmological constant. These dimensionful parameters may take, for reasons not understood, special values, rather than their ``natural values" which are of order the Planck scale. Once the scalar masses are set at these special (or fine-tuned) values, we assume that the corrections to $\mu_\phi^2$ arising from other particles present in the model do not exceed the physical mass of $\phi$. 

A positive mass parameter turning negative via RGE flow leads to dimensional transmutation as can be seen in a Coleman-Weinberg \cite{Coleman:1973jx} analysis of the effective potential. The RGE evolution that we employ is in one to one correspondence with the effective potential, where the minimization is performed at a momentum scale close to the mass of the Higgs scalar.

The outline of the paper is as follows. In Sec. \ref{sec:SM and type-I}, we discuss the absence of such radiative EWSB in SM and type-I seesaw model. Even though such a mechanism fails for type-I seesaw models, in Sec. \ref{sec:Type-II Seesaw} we show that the presence of a TeV scale weak triplet makes radiative EWSB a success for the case of type-II seesaw models. In Sec. \ref{sec:Two-loop neutrino mass model}, we show that for a two-loop neutrino mass model positive Higgs mass parameter at a high energy scale turns negative in the renormalization group flow to low energy. In Sec. \ref{sec: IDM} and Sec. \ref{sec:scalar singlet DM model}, we show that simple dark matter models such as inert doublet model and scalar singlet dark matter model also exhibit radiative EWSB. when the models have TeV scale scalar(s) coupled to the SM Higgs boson. In Sec. \ref{sec:Universal seesaw}, we study the radiative EWSB for a universal seesaw model. Finally in Sec. \ref{sec:conclusion} we conclude.

\section{Absence of radiative EWSB in SM and type-I seesaw models}
\label{sec:SM and type-I}
The mechanism of radiative EWSB occurs when the renormalization group flow of Higgs mass parameter $(\mu_\phi^2)$ receives enough negative contribution from various parameters of the model turning the positive quantity into a negative one while evolving from high to low energies. Unfortunately such cannot be the case in the SM.  

The Higgs potential of the SM is given by:
\begin{equation}
V(\phi)=\mu_\phi^2 \phi^\dagger \phi+\dfrac{\lambda}{2}(\phi^\dagger \phi)^2.\nonumber
\end{equation}
And the renormalization group equation (RGE) of mass parameter $(\mu_\phi^2)$ is given by \cite{SMRGE}:
\begin{equation}
16 \pi^2 \dfrac{d \mu_\phi^2}{dt}=\mu_\phi^2 \left(6 \lambda +2 Tr(3 \mathbf{Y_u^\dagger Y_u}+3 \mathbf{Y_d^\dagger Y_d}+3 \mathbf{Y_e^\dagger Y_e})-\dfrac{9}{10}g_1^2-\dfrac{9}{2}g_2^2\right). \nonumber
\end{equation}
The evolution of the Higgs mass parameter $(\mu_\phi^2)$ is dominated by the gauge couplings and the top quark Yukawa couplings. However, these corrections are proportional to $\mu_\phi^2$ itself, which implies that a positive $\mu_\phi^2$ cannot turn into negative $\mu_\phi^2$ in RGE evolution making radiative EWSB an impossibility within SM.

For type-I seesaw models, the additional Lagrangian is given by:
\begin{equation}
\mathcal{L} \supset -(\mathbf{Y_{\bm{\nu}}})_{ij}\; v \overline{\nu_R}^i \nu_L^j-(\mathbf{M_R})_{ij}\;\overline{\nu_R}^i \text{C}\; \nu^j_R. \nonumber
\end{equation}
This part of the Lagrangian manages to contribute in the renormalization group flow as it adds a new term to the RGE of Higgs mass parameter:
\begin{equation}
16 \pi^2 \dfrac{d \mu_\phi^2}{dt}=16 \pi^2 \left(\dfrac{d \mu_\phi^2}{dt}\right)_{SM}-4\; \tr (\mathbf{Y_\nu Y_\nu^\dagger M_R^\dagger M_R})\nonumber
\end{equation}
Unfortunately, the contributions coming from the $\nu_R$ fields make the situation worse as they only strengthen the positivity of the mass parameter as it evolves from high to low energies. One should also notice that if we want to use the criterion of naturalness, ie. the correction to the Higgs mass parameter $\lesssim 1 \; \text{TeV}^2$ the scale of the Majorana mass of the right-handed neutrino should not exceed $7.4 \times 10^7 \; \text{GeV}$ \cite{Vissani:1997ys}.

In supersymmetry, the radiative EWSB becomes successful \cite{SUSYREWSB} as the scalar partners of the fermions contribute significantly in the renormalization group flow of the Higgs mass parameter in the right direction, making the positive term negative as it evolves from high to low energies. Similar incidents occur in other extensions of SM such as type-II seesaw models where TeV scale particle(s) manages to dominate the renormalization group flow and turn the positive value of the Higgs mass parameter into negative value triggering radiative EWSB. A common feature of these extensions is the presence of new scalar(s) at the TeV scale as we show in the next five sections.

\section{Radiative EWSB in a type-II seesaw neutrino mass model}
\label{sec:Type-II Seesaw}
While type-I seesaw needs right-handed (RH) neutrinos which are neutral under the Standard Model gauge group with large Majorana masses, the minimal type-II seesaw mechanism requires the existence of a weak scalar triplet. The most natural source for such triplets is provided by the Left-Right symmetric theories which can be realized either at low energy or can be embedded in Grand Unified Theories (GUTs) such as $SO(10)$ or $E_6$. Here we study the potential radiative EWSB scenario of the type-II seesaw extension of the SM. 
\begin{figure}[htb]
\begin{center}
\includegraphics[width=5cm]{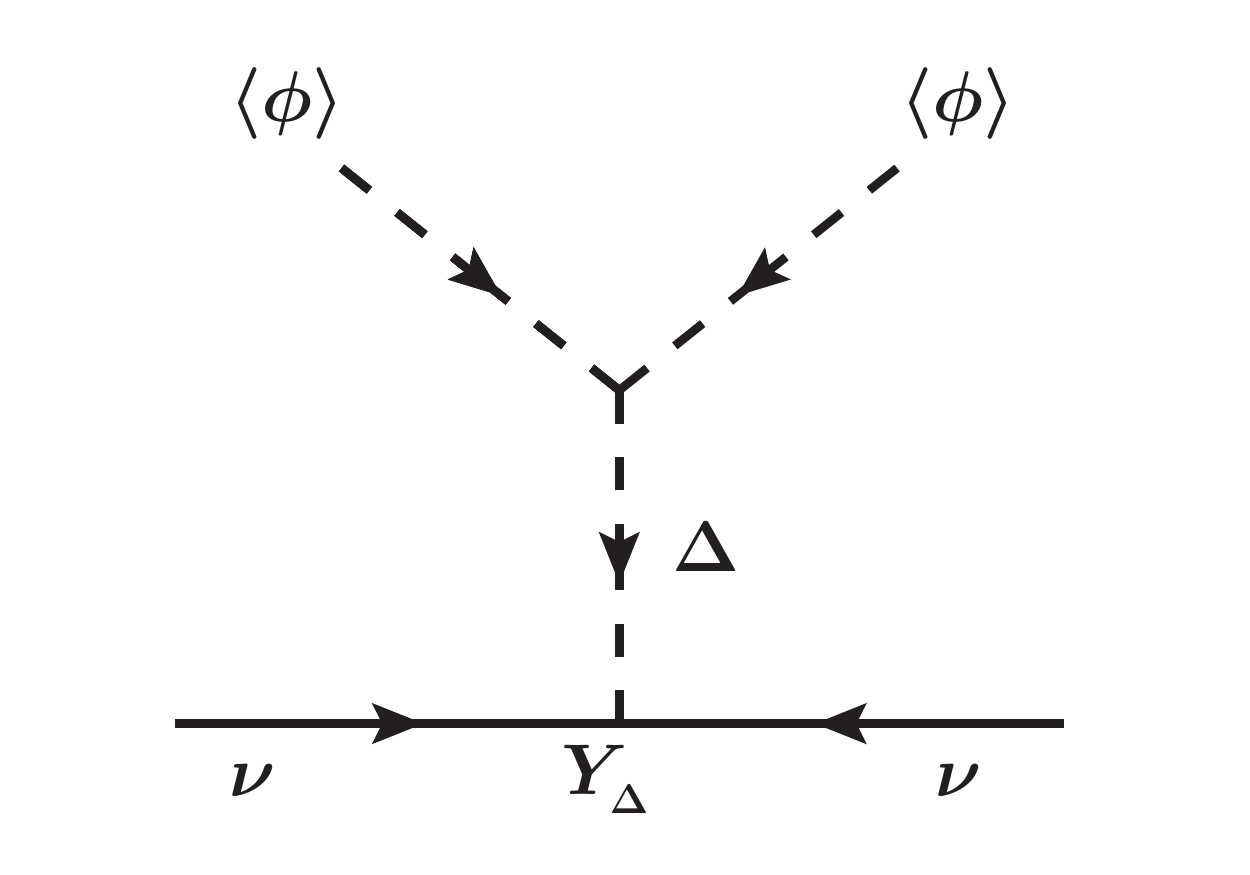}
\caption{Diagrammatic representation of type-II seesaw.}
\end{center}
\end{figure}
\subsection{The model}
We consider the possibility that the weak scalar triplet $\Delta (1,3,1)$ is the only low-energy remnant of the new physics beyond the SM and the neutral component $(\Delta^0)$ acquires a very small induced VEV at low energy. The SM electroweak doublet $\phi (1,2,\nicefrac{1}{2})$ and the electroweak triplet $\Delta ( 1,3,1)$ are denoted by:
\begin{equation}
\phi=\begin{pmatrix}
\phi^+\\
\phi^0
\end{pmatrix}; \hspace{1cm} \Delta = \dfrac{\sigma_i}{\sqrt{2}}\;\Delta_i= \begin{pmatrix}
\nicefrac{\Delta^+}{\sqrt{2}} & \Delta^{++} \\
\Delta^0 & \nicefrac{\Delta^+}{\sqrt{2}}
\end{pmatrix} 
\end{equation}
where $\sigma_i$'s are the Pauli matrices. The most general renormalizable tree-level scalar potential is
\begin{equation}
\label{eq:type-II scalar potential}
\begin{split}
V(\phi,\Delta)=&\; \mu_\phi^2 \; \phi^\dagger \phi +\dfrac{\lambda_1}{2}\; (\phi^\dagger \phi)^2+ \mu_\Delta^2 \; \tr ( \Delta^\dagger \Delta)+\dfrac{\lambda_2}{2} \l(\tr ( \Delta^\dagger \Delta)\r)^2 +\dfrac{\lambda_3}{2} \[ \l(\tr ( \Delta^\dagger \Delta)\r)^2- \tr (\Delta^\dagger \Delta \Delta^\dagger \Delta) \]\\ 
& + \lambda_4 \; \phi^\dagger \phi \; \tr ( \Delta^\dagger \Delta)+ \lambda_5 \; \phi^\dagger [\Delta^\dagger , \Delta ] \phi +\{ \dfrac{\mu}{\sqrt{2}}\; \phi^T\; i \sigma_2 \; \Delta^\dagger \phi +h.c.\}.
\end{split}
\end{equation}
The weak triplet also generates a Majorana mass term for the neutrinos through the Yukawa Lagrangian:
\begin{equation}
\label{eq:type II Yukawa}
\mathcal{L}_Y \supset - \dfrac{(\mathbf{Y_\Delta})_{ij}}{\sqrt{2}}\;\ell_i^T C \; i \sigma_2 \; \Delta \; \ell_j +h.c.
\end{equation}
With the VEV of the electroweak doublet $\langle \phi \rangle =v$, an effective dimension 5 operator generates the neutrino masses through the small but non-zero induced VEV, $\langle \Delta \rangle=\dfrac{\mu v^2}{\sqrt{2}\mu_\Delta^2} \ll v \; \text{when } \;\;v \ll \mu_\Delta\;\; $ and /or  $\mu$ is so small that $\mu v^2 \ll \mu_\Delta^2$ as the electroweak triplet decouples. The neutrino mass matrix is given by:
\begin{equation}
\label{mass neutrino}
\mathbf{m_\nu} \simeq \;\mathbf{Y_\Delta} \;\dfrac{ \mu \; v^2}{2 \mu_\Delta^2}.
\end{equation}
Here one assumes $\mu_\Delta^2>0$ so that $ \langle \Delta \rangle$ is induced only after $ \langle \phi \rangle =v$ is generated.

One also needs to realize the fact that integrating out the heavy scalar triplet in the tree level approximation will also have an effect on the SM Higgs quartic coupling. The effective quartic coupling below the scale $ \mu_r=\mu_\Delta$ is given by:
 \begin{equation}
 \label{quartic matching condition}
 \lambda_1^{\text{eff}}=\lambda_1-\dfrac{\mu^2}{\;\mu_\Delta^2}.
 \end{equation}
This is the connecting formula for the Standard Model quartic coupling $\lambda_1$ at the scale $ \mu_r=\mu_\Delta$.
 \subsection{The stability conditions for the Higgs potential and the evolution of mass parameters}
One needs to be careful while studying the solution to the set of RGEs of the parameters of a model. The parameters in the scalar potential have to satisfy certain conditions at all energy scales which ensures that the potential is bounded from below. For that purpose one must identify the necessary and sufficient conditions for boundedness of the potential. For type-II seesaw models, the stability conditions for the potential to be bounded from below can be derived to be:
\begin{enumerate}[label=(\roman*)]
\item \hfill  \makebox[5pt][r]{%
            \begin{minipage}[b]{\textwidth}
              \begin{equation}
              \label{cond1}
                 \lambda_1 \geq 0 \; ; \hspace{.2cm} \lambda_2 \geq 0\; ; \hspace{10cm}
              \end{equation}
          \end{minipage}}
 \item \hfill  \makebox[5pt][r]{%
            \begin{minipage}[b]{\textwidth}
              \begin{equation}
\label{cond2}                
                 \lambda_4  -\left| \lambda_5 \right| \geq-\sqrt{\lambda_1 \lambda_2} \; ; \hspace{.2cm} 2 \lambda_2+\lambda_3 \geq 0; \hspace{6.7cm}
              \end{equation}
          \end{minipage}}
 \item
  \hfill  \makebox[5pt][r]{%
            \begin{minipage}[b]{\textwidth}
              \begin{equation}
                 2 \lambda_4 \sqrt{\lambda_2}+2 \lambda_2 \sqrt{\lambda_1} +\lambda_3 \; \sqrt{\lambda_1} \geq 0 \;  \hspace{8.1cm} \nonumber 
              \end{equation}
          \end{minipage}}
                      \begin{equation}
\label{cond3}
              \text{or,} -2 \lambda_1 \lambda_2 \lambda_3 -\lambda_1 \lambda_3^2+2 \lambda_3 \lambda_4^2-2\left(2 \lambda_2 +\lambda_3\right) \lambda_5^2 \geq 0\; ; \hspace{4cm}
              \end{equation}
\end{enumerate}
This is compact set of constraints that is necessary and sufficient as we show in the Appendix \ref{appendix: stability cond}. For a less compact set of constraints see Ref. \cite{Arhrib:2011uy}. The couplings of the Lagrangian have to maintain these stability conditions upto the energy scale of new physics such as GUTs.

Using vertex corrections and the wave function renormalization factors, we can calculate the complete set of $\beta$-functions and renormalization group equations (RGEs). We have also determined the RGEs for the mass parameters of the model which are related to the anomalous dimensions $(\gamma_m)$ of the scalar masses by 
 \begin{equation}
 \gamma_m \equiv \dfrac{1}{2}\dfrac{d \ln(m^2)}{d t}
 \end{equation}
where $t=\ln \mu_r$ and $\mu_r$ is the running scale. The set of RGEs for the mass parameters is given by \footnote{We disagree with the signs of terms involving the couplings $\lambda_4$ and $\lambda_5$ in the RGE of the mass parameter $\mu$ given in Eq. (17) of Ref. \cite{Schmidt:2007nq}. We also disagree with the coefficient in front of the mass parameter $|\mu|$ in the RGE of $\mu^2_\Delta$ given in Eq. (18) of Ref. \cite{Schmidt:2007nq}.}:
\begin{equation}
\label{mass RGEs Type-II}
\begin{split}
16\pi^2 \dfrac{d \mu_\phi^2}{d t} =&\[- \frac{9}{10}g_1^2-\frac{9}{2}g_2^2+6\lambda_1+2 T\] \mu^2_\phi +6 \lambda_4 \mu_\Delta^2+6|\mu|^2;\\
16\pi^2 \dfrac{d \mu_\Delta^2 }{d t}  =&\left[\l(-\frac{18}{5}g_1^2-12g_2^2\r)+8\lambda_2+2\lambda_3+2\tr\left(\mathbf{Y_\Delta^\dagger
Y_\Delta}\right)\right]\mu_\Delta^2+4\lambda_4 \mu_\phi^2+2|\mu|^2;\\
16\pi^2 \dfrac{d \mu }{d t} =&\left[\lambda_1+4\lambda_4-8\lambda_5-\frac{27}{10}g_1^2-\frac{21}{2}g_2^2
  +2 T +  \tr\left(\mathbf{Y_\Delta^\dagger Y_\Delta}\right)\right]\mu;\\
\end{split}
\end{equation}
where
\begin{equation}
T =\tr \[\mathbf{Y_e^\dagger Y_e}  +3 \mathbf{Y_d^\dagger Y_d} +3 \mathbf{Y_u^\dagger Y_u} \] .
\end{equation} 

A complete set of RGEs for all the couplings of the Lagrangian is given in the Appendix~\ref{appendix: RGEs for Type-II}. With this set of RGEs we proceed towards its numerical solution. We can already see that the contribution from the cubic coupling $\mu$ and quartic couplings such as $\lambda_4$ has the potential to drive the positive mass parameter $\mu_\phi^2$ at high energy scale towards a negative value at lower energy scale to trigger Radiative EWSB. For a TeV scale scalar triplet mass one realizes that to get the correct order of neutrino mass the cubic parameter $\mu$ needs to be very small  ($\sim~10^{-5}$~GeV), which makes the contribution of $\mu$ term in the RGEs of mass parameters irrelevant. This choice is natural, since quantum corrections to $\mu$  are proportional to $\mu$ itself - see Eq. (\ref{mass RGEs Type-II}). 
\subsection{Solution to the RGEs}
To analyze the evolution of the mass parameters, one needs to solve the set of RGEs which in turn requires one to define the relevant couplings at some energy scale \cite{typeIIphen}. In this case, all the SM gauge couplings, Yukawa couplings and Higgs quartic couplings were evaluated at two-loop level upto the energy scale corresponding to the scalar triplet mass. At energies above the triplet mass the gauge couplings were evolved continuously but with the updated RGEs given in the set of RGEs in Eq. (\ref{RGEs Type-II gauge}). The quartic coupling of the SM electroweak doublet Higgs has a discontinuity at the triplet mass scale due to the matching condition of the parameter given in Eq. (\ref{quartic matching condition}). Above the energy scale $\mu_\Delta$, the full set of RGEs was used to evolve all the parameters of the model. 
\begin{table}[!h]
\centering
\begin{center}\renewcommand\arraystretch{1.3}
\begin{tabular}{||c|c||c|c||}   
 \hline 
  \hline 
 Quartic couplings & values &  Mass parameters & values\\ 
 \hline 
  \hline 
  $\lambda_1(m_Z)$ & $0.258$&$m_t(m_t)$ & $162.25$ GeV\\ 
  $\lambda_1(\mu_\Delta)$ & $0.1887$&$\mu_\Delta^2(\mu_\Delta)$ & $500^2\; \text{(GeV)}^2$\\  
  $\lambda_2(\mu_\Delta)$&$0.15$&$v(m_Z)$ & $174.10$ GeV \\
  $\lambda_3(\mu_\Delta)$&$0.45$&$\mu^2_\phi(125 \; \text{GeV})$& $-(88.91)^2 \; \text{(GeV)}^2$\\
$\lambda_4(\mu_\Delta)$&$0.19$&$\mu^2_\phi(\mu_\Delta)$& $-(89.59)^2\; \text{(GeV)}^2$\\
$\lambda_5(\mu_\Delta)$&$0.10$&$\mu(\mu_\Delta)$&$10^{-5}$ GeV\\
 \hline
  \hline 
 \end{tabular}  
 \end{center}
\caption{Quartic couplings and mass parameter values for the sample point used for the type-II seesaw model in Fig. \ref{fig:sample RGE Type-II seesaw}.}
\label{table: sample point Type-II seesaw}
\end{table}

\begin{figure}[!h]
\centering
\begin{subfigure}{.5\textwidth}
  \centering
  \includegraphics[width=8cm]{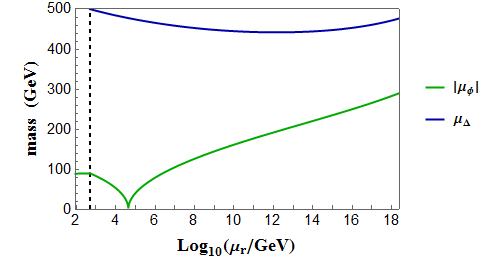}
  \caption{Evolution of mass parameters}
  \label{fig:sub1tII}
\end{subfigure}%
\begin{subfigure}{.5\textwidth}
  \centering
  \includegraphics[width=8cm]{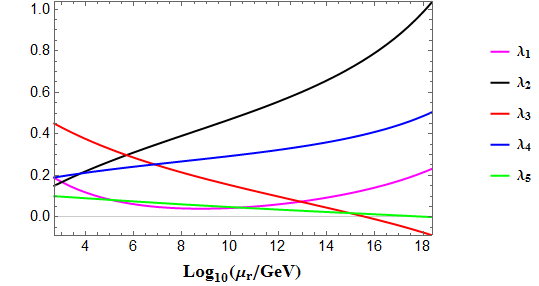}
  \caption{Evolution of quartic couplings}
  \label{fig:sub2tII}
\end{subfigure}
\caption{One-loop running of the parameters of type-II seesaw model from Planck scale down to weak scale. The black dashed line in Fig. \ref{fig:sub1tII} corresponds to the scale, $\mu_r=\mu_\Delta$. In Fig. \ref{fig:sub1tII} the evolution of the absolute value of the SM Higgs mass parameter $\left(\left| \mu_\phi\right|\right)$ along with the mass of the weak triplet scalar $\left(\mu_\Delta\right)$ has been plotted. The point at which $\left| \mu_\phi\right|$ touches the horizontal axis corresponding to mass =$\;0 \; \text{GeV}$, is the energy scale where radiative EWSB is triggered as the sign of the SM Higgs mass-squared parameter $( \mu_\phi^2)$ switches from positive to negative while evolving from high to low energies. Note that in this sample case, the radiative EWSB is prompted at around an energy scale of $30\; \text{TeV}$. Fig. \ref{fig:sub2tII} shows the evolution of all the quartic couplings of the type-II seesaw model from Planck scale down to weak scale emphasizing the fact that the model remains perturbative all the way for the selected sample point.}
\label{fig:sample RGE Type-II seesaw}
\end{figure}
To generate a sample case, we specified the values of the quartic couplings and mass parameters of the model at the low energy scale, $\mu_r=\mu_\Delta$, consistent with the stability conditions given in the inequalities Eqs. (\ref{cond1}), (\ref{cond2}), and (\ref{cond3}). Also the  masses of neutrinos put a natural limit on the cubic coupling of the model because of Eq. (\ref{mass neutrino}) and this in turn makes the discontinuity in the Higgs quartic coupling $\lambda_1$ ignorable. To illustrate the phenomenon of radiative electroweak symmetry breaking in the Type-II seesaw model a sample point is chosen as given in Table \ref{table: sample point Type-II seesaw}. The sample point  satisfies all the stability conditions and the mass parameter $\mu^2_\phi$ runs with a positive slope with the energy scale. Fig. \ref{fig:sample RGE Type-II seesaw} shows that the mass parameter becomes negative at low energy even though it is positive at high energy scale. This turning occurs at $\mu_r \approx 10^5 \; \text{GeV}$, when $|\mu_\phi|$ plotted in the Fig. \ref{fig:sub1tII} becomes zero. This analysis shows that radiative electroweak symmetry breaking may be successfully achieved in type-II seesaw models. The mass of the triplet scalar should remain below about a few TeV or else $\mu_\phi^2$ becomes too negative.

\section{Radiative EWSB in a two-loop  neutrino mass model}
\label{sec:Two-loop neutrino mass model}
Even before the experimental discovery of neutrino oscillation which is a clear indication of non-zero neutrino masses and mixings, the subject of neutrino mass generation has been an active arena of research. An interesting alternative to the type-I or type-II seesaw mechanism is that the neutrino masses are generated by loop corrections, hence the masses are suppressed by the loop factors. In this scenario, the new particles responsible for the neutrino mass generation should be relatively light with the possibility that they show up at the Large Hadron Collider in the near future.
\subsection{The Model}
We will consider the two-loop neutrino mass model which introduces a doubly charged $(k^{++})$ and a singly charged $(h^+)$ scalars along with the SM particles \cite{Babu:1988ki, Babu:2014kca}. In this model, lepton number is explicitly broken and as a result tiny Majorana mass arises through loop diagram at two-loop level. One of the salient features of the model is that one of the three neutrino masses is very nearly zero. The model admits both normal and inverted hierarchy of neutrino masses and also has CP violation in neutrino oscillations.

The new scalars under the SM gauge group $SU(3)_C\times SU(2)_L \times U(1)_Y$ are denoted by
\begin{equation}
h^+ (1,1,1); \hspace{2cm}k^{++}(1,1,2).
\end{equation}
The gauge invariant Yukawa couplings that are allowed involving the new scalars are:
\begin{equation}
\mathcal{L}_Y \supset \mathbf{f}_{ab}\; \ell_a^i \ell_b^j \epsilon_{ij} h^+ + \mathbf{h}_{ab}\; e^c_a e^c_b k^{--}+h.c.
\end{equation}
Here $a,b$ are generation indices, $i,j$ are $SU(2)_L$ indices with $\epsilon_{ij}$ being antisymmetric tensor. The Yukawa coupling matrices $\mathbf{f}$ and $\mathbf{h}$ are antisymmetric and symmetric respectively. 

\begin{figure}[htb]
\begin{center}
\includegraphics[width=9cm]{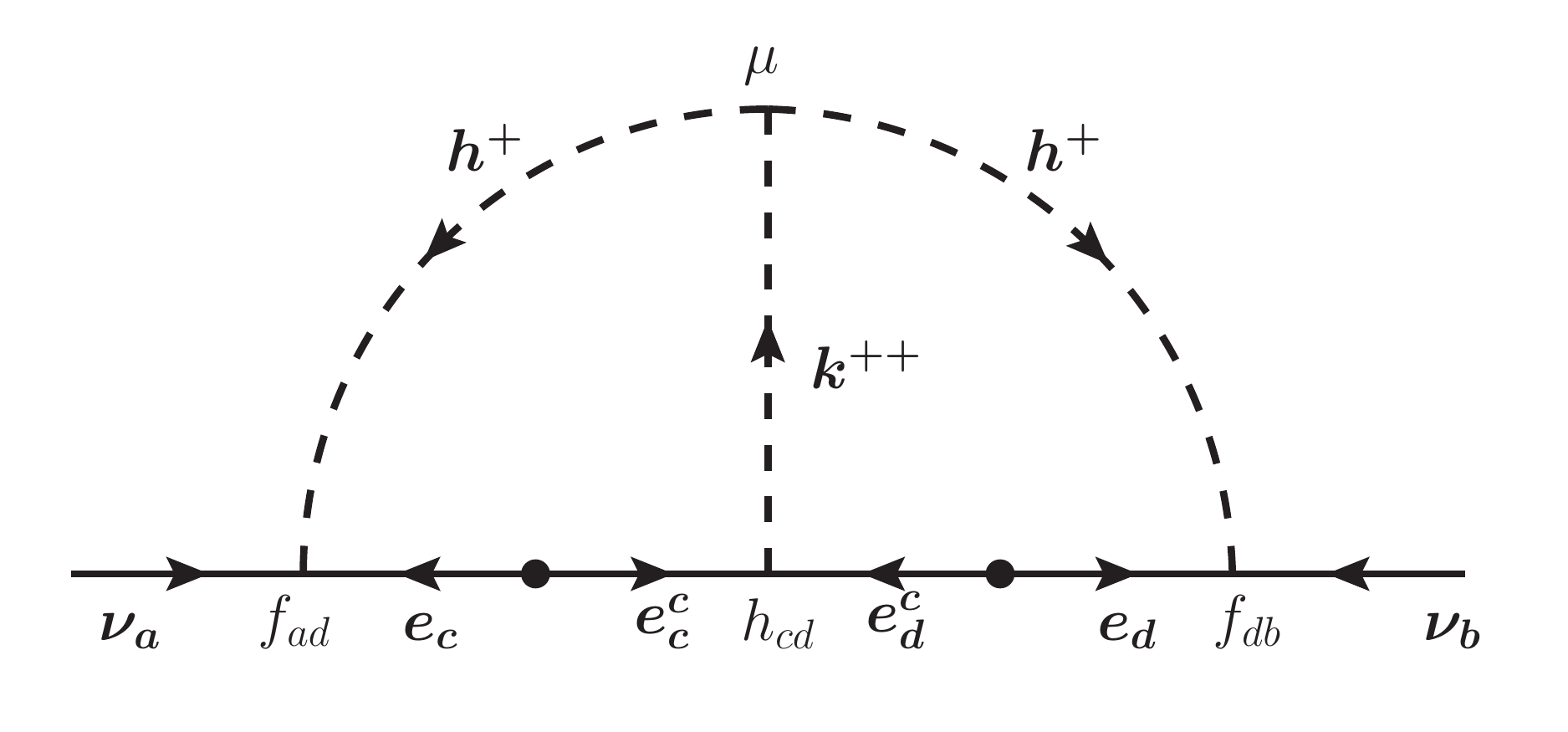}
\caption{Feynman Diagram responsible for neutrino mass generation at two-loop level.}
\label{fig:Two-loop neutrino dia}
\end{center}
\end{figure}

The scalar potential for the model is given by:
\begin{equation}
\begin{split}
V(\phi, h^+,k^{++})&= \mu_\phi^2 \; \phi^\dagger \phi + \mu_h^2 \;h^+ h^-+\mu_k^2 \;k^{++} k^{--} -\l(\mu \;h^+h^+k^{--}
+h.c. \r) +\dfrac{\lambda_1}{2}\;  (\phi^\dagger \phi)^2\\ & + \dfrac{\lambda_2}{2}\;  (h^+ h^-)^2+\dfrac{\lambda_3}{2}\;  (k^{++}k^{--})^2+\lambda_4\;(\phi^\dagger \phi)(h^+h^-)+\lambda_5\; (\phi^\dagger \phi) (k^{++}k^{--})\\ &+ \lambda_6 \;(h^+h^-)(k^{++}k^{--}).
\end{split}
\end{equation}
Small neutrino masses are generated matrix is generated by a two-loop process involving the couplings $\mathbf{f}$, $\mathbf{h}$ and $\mu$ $-$ the simultaneous presence of these couplings break lepton number $-$ depicted in the Feynman diagram shown in Fig. \ref{fig:Two-loop neutrino dia}.
Neutrino oscillation phenomenology of this model has been studied extensively \cite{Babu:2002uu, twoloop, Babu:2014kca}. Our goal here is to study the renormalization group flow of the Higgs mass parameter and determine the possibility of radiative EWSB. While performing such a study, one also needs to be certain that the scalar potential of the model remains bounded and also that the model remains perturbative. 
\subsection{The stability conditions for the Higgs potential and the evolution of mass parameters}
The necessary and sufficient boundedness conditions for the Higgs potential which ensures that the potential is bounded from below are given by \cite{Babu:2014kca}:
\begin{equation}
\label{eq:stability conditions two-loop}
\begin{split}
\text{(i)} & \; \lambda_1 \geq 0; \hspace{1cm} \lambda_2 \geq 0; \hspace{1cm} \lambda_3 \geq 0; \\
\text{(ii)}& \; \lambda_4\geq -\sqrt{\lambda_1 \lambda_2} ;\hspace{1cm} \lambda_5\geq -\sqrt{\lambda_1 \lambda_3} ;\hspace{1cm} \lambda_6\geq -\sqrt{\lambda_2 \lambda_3} ;\hspace{1cm} \\
\text{(iii)}& \; \lambda_4\sqrt{\lambda_3}+\lambda_6\sqrt{\lambda_1}+\lambda_5\sqrt{\lambda_2}+\sqrt{\lambda_1 \lambda_2 \lambda_3}\geq 0\hspace{1cm} \text{or} \hspace{1cm} \det{\bm{\lambda}} \geq 0;
\end{split}
\end{equation}
where
\begin{equation}
\bm{\lambda}=\begin{pmatrix}
\lambda_1 & \lambda_4 &\lambda_5\\
\lambda_4 & \lambda_2 &\lambda_6\\
\lambda_5 & \lambda_6 &\lambda_3\\
\end{pmatrix}.
\end{equation}
It has been shown that the model maintains perturbativity all the way to the Planck scale and boundedness for both normal and inverted case, if $|\mathbf{h}_{\mu\mu}|<0.45$ and $|\mathbf{f}_{\mu \tau}|<~0.34$ \cite{Babu:2014kca}. For the antisymmetric Yukawa coupling matrix $\mathbf{f}$, the neutrino mixing angles provide two constraints, reducing the number of free parameters to one. For the case of normal neutrino mass hierarchy the relation is given by \cite{Babu:2002uu}:
\begin{equation}
\label{eq:yukawa f relation}
\begin{split}
\epsilon =& \tan \theta_{12}\; \dfrac{\cos\theta_{23}}{\cos\theta_{13}}+\tan\theta_{13}\;\sin\theta_{23} \;e^{-i \delta};\\
\epsilon' =& \tan \theta_{12}\; \dfrac{\sin\theta_{23}}{\cos\theta_{13}}-\tan\theta_{13}\;\cos\theta_{23} \;e^{-i \delta}.\\
\end{split}
\end{equation}
And for the inverted mass hierarchy we have:
\begin{equation}
\epsilon=-\sin\theta_{23}\; \cot\theta_{13}\;e^{-i\delta}; \hspace{1cm} \epsilon'= \cos\theta_{23}\;\cot\theta_{13}\;e^{-i\delta}.
\end{equation}
where in both case $\epsilon$ and $\epsilon'$ are defined as:
\begin{equation}
\label{eq:yukawa f def}
\epsilon \equiv \dfrac{\mathbf{f}_{e\tau}}{\mathbf{f}_{\mu\tau}}; \hspace{1cm} \epsilon' \equiv \dfrac{\mathbf{f}_{e\mu}}{\mathbf{f}_{\mu\tau}}.
\end{equation}

Similar to the analysis of type-II seesaw model in Sec. \ref{sec:Type-II Seesaw}, we require the full set of RGEs for this model which includes the evolution of the gauge couplings, Yukawa couplings, quartic couplings and the mass parameters of the Lagrangian. While the complete set of the RGEs is listed in the Appendix \ref{appendix: RGEs for Two-loop}, the RGEs for the mass parameters of the Lagrangian are given by:
\begin{equation}
\label{mass RGEs two-loop neutrino}
\begin{split}
16 \pi^2 \dfrac{d \mu_\phi^2}{d t} =& \mu_\phi^2 \l( -\dfrac{9}{10} g_1^2 -\dfrac{9}{2} g_2^2+2 T +6 \lambda_1\r) +2 \lambda_4 \mu_h^2+2 \lambda_5 \mu^2_k;\\
16 \pi^2 \dfrac{d\mu_h^2}{d t} =& \mu_h^2 \l( -\dfrac{18}{5} g_1^2 +8 \tr(\mathbf{f^\dagger f})  +4 \lambda_2\r) +4 \lambda_4 \mu_\phi^2+2 \lambda_6 \mu^2_k+8 \mu^2;\\
16 \pi^2 \dfrac{d\mu_k^2}{d t} =& \mu_k^2 \l( -\dfrac{72}{5} g_1^2 +4 \tr(\mathbf{h^\dagger h}) +4 \lambda_3\r) +4 \lambda_5 \mu_\phi^2+2 \lambda_6 \mu^2_h+4 \mu^2;\\
16 \pi^2 \dfrac{d\mu}{d t}=& \mu \l( -\dfrac{54}{5}g_1^2+2\lambda_2+2\lambda_6+2 \tr(\mathbf{h^\dagger h})+8 \tr(\mathbf{f^\dagger f})  \r). \\
\end{split}
\end{equation}
Note that the SM Higgs mass parameter $(\mu_\phi^2)$ can be turned negative proportional to $\mu_h^2$ and/or $\mu_k^2$ in going from high energy to low energy as long as either $\lambda_4$ or $\lambda_5$ is positive. Such a choice is consistent with the boundedness conditions given in Eq. (\ref{eq:stability conditions two-loop}), thus enabling radiative EWSB within the model.
\subsection{Solution to the RGEs}
To find the solution to the full set of RGEs, one requires to completely specify the values of all the parameters of the Lagrangian at some energy scale. We specify the sample values at low energy scale while satisfying the necessary and sufficient conditions for the boundedness of the scalar potential in Table \ref{table: sample point two-loop model}. 
\begin{table}[!ht]
\centering
\begin{center}\renewcommand\arraystretch{1.3}
\begin{tabular}{||c|c||c|c||} 
 \hline 
  \hline 
Quartic, Yukawa couplings & values &  Mass paramters & values\\ 
 \hline 
  \hline 
$\lambda_1(M_Z)$ & $0.258$ & $m_t(m_t)$ & $162.25$ GeV\\ 
$\lambda_1(\mu_0)$ & $0.1924$&$M_h(m_Z)$ & $125.1$ GeV\\ 
 $\lambda_2(\mu_0)$&$0.20$&$v(m_Z)$ & $174.10$ GeV \\
 $\lambda_3(\mu_0)$&$0.50$&$\mu(\mu_0)$&$500$GeV\\
$\lambda_4(\mu_0)$&$0.10$&$\mu_h^2(\mu_0)$ & $800^2 \; \text{(GeV)}^2$ \\
$\lambda_5(\mu_0)$&$0.15$&$\mu_k^2(\mu_0)$ & $450^2 \; \text{(GeV)}^2$\\
$\lambda_6(\mu_0)$&$-0.25$ &$\mu^2_\phi(125 \; \text{GeV})$ &$-(88.91)^2 \; \text{(GeV)}^2$\\
$|\mathbf{f_{\mu \tau}}| (\mu_0)$&$0.013$&$\mu_\phi^2(\mu_0)$ &$-(89.55)^2 \; \text{(GeV)}^2$\\
$|\mathbf{h_{\mu \mu}}| (\mu_0)$&$0.4$&\\
\hline
  \hline 
 \end{tabular}  
 \end{center}
\caption{Quartic and Yukawa coupling and mass parameter values for the sample point used for the Two-loop neutrino mass model in Fig. \ref{fig:sample RGE Two-loop neutrino mass model}}
\label{table: sample point two-loop model}
\end{table}

For the sample case, we selected the normal mass hierarchy for no specific reason. Similar result can be found if the hierarchy is inverted. We used the set of two-loop RGEs for the SM case and ran upto the lightest newly introduced scalar particle (in the sample point $\mu_0=\mu_k$). The full set of RGEs  (given in Appendix \ref{appendix: RGEs for Two-loop} along with Eq. (\ref{mass RGEs two-loop neutrino})) was used to evolve the couplings and the mass parameters from the energy scale $\mu_0$ upto the Planck scale. 

In Table \ref{table: sample point two-loop model} only the value of $|\mathbf{f_{\mu \tau}}|$ is listed as the other values of the Yukawa couplings $\mathbf{f}$ can be calculated using Eq.~(\ref{eq:yukawa f relation}) and the definitions in Eq. (\ref{eq:yukawa f def}).  For the case of the Yukawa couplings $\mathbf{h}$, we only kept the value of $|\mathbf{h_{\mu\mu}}|$ non-zero as in the fit to neutrino masses within this model, $\left|\mathbf{h_{\mu \mu}}\right| \gg \left| \mathbf{h}_{ij} \right|$ for all $i,j \neq 2,2$ \cite{Babu:2014kca}.

\begin{figure}[!tb]
\centering
\begin{subfigure}{.55\textwidth}
  \centering
  \includegraphics[width=8cm]{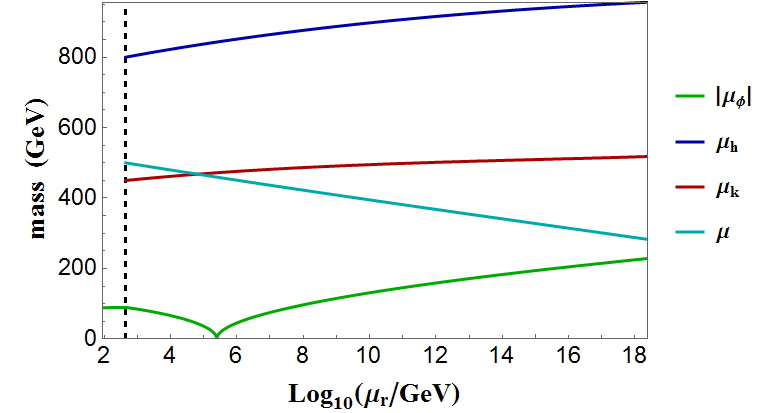}
  \caption{Evolution of mass parameters}
  \label{fig:sub1twolopp}
\end{subfigure}%
\begin{subfigure}{.45\textwidth}
  \centering
  \includegraphics[width=8cm]{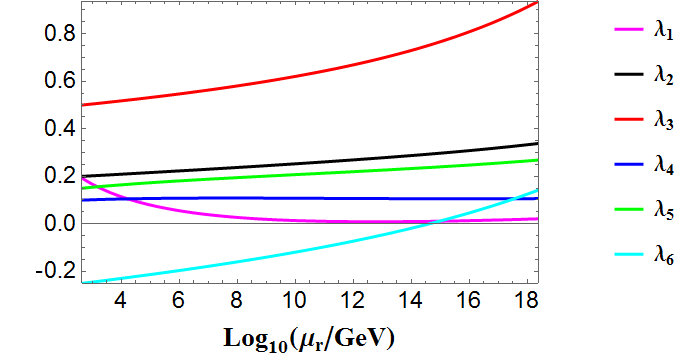}
  \caption{Evolution of quartic couplings}
  \label{fig:sub2twoloop}
\end{subfigure}
\caption{One-loop running of the parameters of two-loop neutrino mass model from Planck scale down to weak scale. The black dashed line corresponds to the scale, $\mu_r=\mu_0$. Here $\mu_0$ is the energy scale corresponding to the lightest of the newly introduced particles. In this sample point $\mu_o=\mu_k$. In Fig. \ref{fig:sub1twolopp} the evolution of the absolute value of the SM Higgs mass parameter $\left(\left| \mu_\phi \right| \right)$ along with $\mu_h$, $\mu_k$ and cubic coupling $\mu$ has been plotted. The point at which $\left| \mu_\phi \right|$ touches the horizontal line corresponding to mass =$\;0 \; \text{GeV}$, is the energy scale where radiative EWSB is triggered as the sign of the SM Higgs mass-squared parameter $( \mu_\phi^2)$ switches from positive to negative while evolving from high to low energies. Note that $\mu_\phi^2$ turns negative around $\mu_r \approx 10^5 \; \text{GeV}$, while $\mu_k^2$  and $\mu_h^2$ remain positive. Fig. \ref{fig:sub2twoloop} shows the evolution of all the quartic couplings of the two-loop neutrino mass model from Planck scale down to weak scale emphasizing the fact that the model remains perturbative all the way for the selected sample point.}
\label{fig:sample RGE Two-loop neutrino mass model}
\end{figure}

Upon performing a numerical analysis to solve the full set of RGEs for the model, we plot the evolution of all the mass parameters and the quartic couplings of the model in Fig. \ref{fig:sample RGE Two-loop neutrino mass model}. The sample point satisfies all the necessary and sufficient boundedness conditions given in Eq. (\ref{eq:stability conditions two-loop}) and Fig. \ref{fig:sub2twoloop} ensures that all the quartic couplings stays within the perturbative range up to Planck scale.   As we have plotted the absolute value of the SM Higgs mass parameter $(\left| \mu_\phi \right|)$, the point at which the plot of $\left| \mu_\phi \right|$ touches the horizontal axis (ie. mass $=\; 0\; \text{GeV}$) corresponds to the energy scale where the positive mass-squared parameter $(\mu_\phi^2)$ turns negative at low energy, triggering radiative EWSB. For the selected sample point, the radiative correction manages to push the Higgs mass parameter in such a way that it acquires a negative value at the energy scale $\mu_r \approx 10^5 \; \text{GeV}$.


\section{Inert doublet model}
\label{sec: IDM}
Inert doublet model is one of the simplest extensions of the SM, which can be treated as a special case of more general two Higgs doublet model. In this model, the Lagrangian has a $\mathbb{Z}_2$ symmetry that remains unbroken by the vacuum structure. Even though it was introduced in the 70's \cite{Deshpande:1977rw}, it received a new influx of attention when the model was shown to be able to alleviate the issue of nondiscovery of Higgs boson up to a mass of $115\;\text{GeV}$ \cite{LEPparadox}, be able to address the issue of the smallness of the neutrino masses either via type-I seesaw mechanism or via one loop radiative mechanism (in a version referred to as the Scotogenic model) \cite{Ma:2006km}, leptogenesis \cite{Ma:2006fn} by including TeV scale right-handed neutrino and most importantly explain dark matter of the universe \cite{Ma:2006km}. It has been shown that electroweak symmetry breaking can be induced by loop effects due to the cross coupling between the SM Higgs and the dark matter candidate of the model \cite{Hambye:2007vf}. We study the scotogenic version of inert doublet model for completeness which contains a dark matter candidate and generates a naturally suppressed neutrino mass at one-loop level. Unlike the general inert doublet model, in the scotogenic version the right-handed neutrinos with sufficiently large masses can via loop effect turn the mass parameter of the inert doublet scalar negative in going from low to high energy; thus breaking the $\mathbb{Z}_2$ symmetry at a high scale. In this situation, the model cannot have a dark matter candidate and the neutrino masses is not naturally suppressed any more \cite{Merle:2015gea}. We consider TeV scale right-handed neutrino masses to avoid such undesired situation.

\subsection{The model}

The scotogenic version of the inert doublet model requires three right-handed neutrinos $(N_i)$ along with inert scalar doublet $(\eta)$ and the SM particles. All the newly introduced particles are charged under the additional $\mathbb{Z}_2$ parity symmetry while all the SM particles are neutral under this parity. The neutral component of the inert scalar doublet is the only dark matter matter candidate for the general inert doublet model while for the scotogenic case the lightest $\mathbb{Z}_2$ odd particle$-$ either a neutral scalar from the doublet, or the right-handed neutrino $-$ is a dark matter candidate. The survival of the $\mathbb{Z}_2$  symmetry is crucial for the model as this symmetry protects the DM candidate from decaying and the same symmetry forbids the neutrinos from acquiring masses at the tree level.

In this model, the right-handed neutrinos get a direct Majorana mass term $\dfrac{1}{2}\overline{N^i_R}M_{ij}N^{j^c}_R+h.c.$ which leads to masses $M_i$'s (where $i=1,2,3$) upon diagonalisation.  As the right-handed neutrinos are odd under the $\mathbb{Z}_2$ symmetry, the neutrino masses cannot be generated at tree level. The Lagrangian contains a neutrino Yukawa coupling involving the inert scalar doublet $\eta$ and the right-handed neutrinos in addition to the SM lepton doublets. This term is given by:
\begin{equation}
\mathcal{L}_Y \supset -\mathbf{h}_{ij}\overline{N^i_R}\tilde{\eta}^\dagger \ell^j_L+h.c.; \hspace{1cm} \text{where}\; \; \tilde{\eta}=i\; \sigma_2 \eta^*
\end{equation}
\begin{figure}[htb]
\begin{center}
\includegraphics[width=5cm]{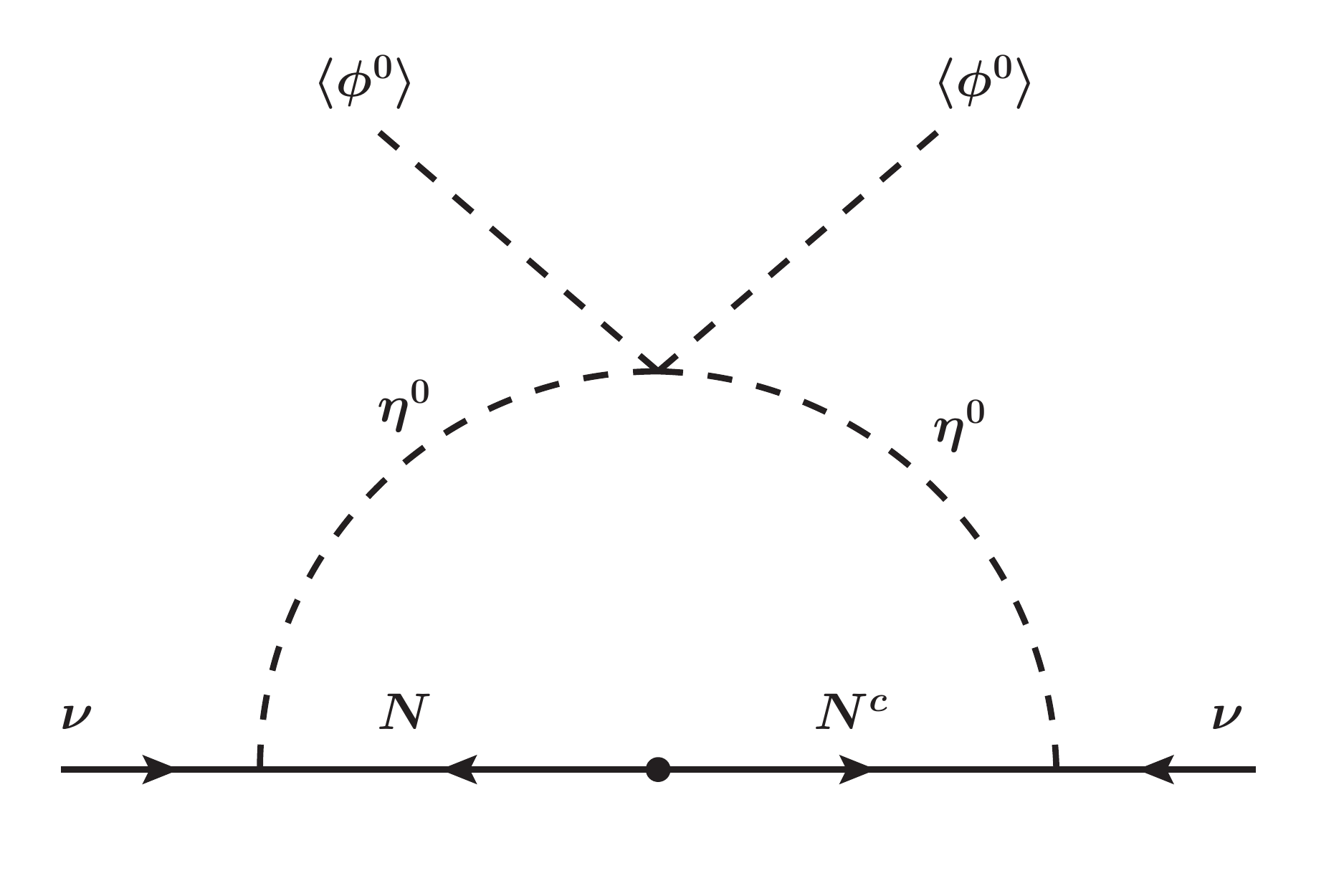}
\caption{Diagrammatic representation of neutrino mass generation in the scotogenic model}
\label{fig:scotogenicmass}
\end{center}
\end{figure}
This additional Yukawa coupling along with the right-handed neutrino Majorana mass term generates the loop suppressed neutrino mass matrix (see Fig. \ref{fig:scotogenicmass}). 

The scalar potential of the model can be written as:
\begin{equation}
\begin{split}
V(\phi, \eta)=&\; \mu_\phi^2 \phi^\dagger \phi + \mu_2^2 \eta^\dagger \eta +\dfrac{\lambda_1}{2} (\phi^\dagger \phi)^2+ \dfrac{\lambda_2}{2} (\eta^\dagger \eta)^2+\lambda_3 (\phi^\dagger\phi)(\eta^\dagger \eta) \\& +\lambda_4(\eta^\dagger \phi)(\phi^\dagger \eta)+\dfrac{\lambda_5}{2}\[(\eta^\dagger \phi)^2+h.c. \]
\end{split}
\end{equation}
In the potential the mass parameters $\mu_i^2 (i=1,2)$ and the couplings $\lambda_i,(i=1-4)$ must be real. $\lambda_5$ can also be taken to real without any loss of generality as the phase of the coupling can be absorbed by the redefinition of the $\eta$ field.

\subsection{The stability conditions and the evolution of mass parameters}
The parameters in the scalar potential have to satisfy the boundedness conditions at all energy scales which ensures that the potential is bounded from below all the way. The conditions are given as:
\begin{equation}
\lambda_1\geq 0; \hspace{.7cm}\lambda_2\geq 0;\hspace{.7cm} \lambda_3\geq -\sqrt{\lambda_1 \lambda_2}; \hspace{.7cm}\lambda_3+\lambda_4-|\lambda_5|\geq -\sqrt{\lambda_1 \lambda_2}.
\end{equation}
One can also find the physical scalar mass spectrum as:
\begin{equation}
\begin{split}
m_h^2=& \;2 \lambda_1 v^2;\\
m_\pm^2=&\;m_2^2+\lambda_3 v^2;\\
m_R^2=& \;m_2^2+v^2 (\lambda_3+\lambda_4+\lambda_5);\\
m_I^2=& \;m_2^2+v^2 (\lambda_3+\lambda_4-\lambda_5).\\
\end{split}
\end{equation}
where $m_h$ is the mass of the SM Higgs boson, $m_\pm$ is the mass of the charged component of $\eta$ doublet, $m_R$ and $m_I$ are respectively the masses of the real and imaginary component of the neutral field inside the $\eta$ doublet. We choose real part of neutral scalar $\eta$ as the DM candidate. For this scenario, the charged component of the electroweak doublet $\eta$ needs to be heavier than the neutral component. Also by keeping $\lambda_5$ negative and small, we get a slightly heavier pseudoscalar.

To illustrate the radiative electroweak symmetry breaking mechanism, we study the RGEs for the mass parameters of the scotogenic version of the inert doublet model which are given by \cite{Merle:2015gea}: 
\begin{equation}
\label{mass RGEs for IDM}
\begin{split}
16 \pi^2 \dfrac{d \mu_\phi^2}{d t}=& 6 \lambda_1 \mu_\phi^2 +2 (2 \lambda_3+\lambda_4)\mu_2^2+\mu_\phi^2 \[2T-\dfrac{3}{2}(g_1^2+3g_2^2)\]; \\
16 \pi^2 \dfrac{d \mu_2^2}{d t}=& 6 \lambda_2 \mu_2^2 +2 (2 \lambda_3+\lambda_4)\mu_\phi^2+\mu_2^2 \[2T_\nu-\dfrac{3}{2}(g_1^2+3g_2^2)\]-4 \sum_{i=1}^3 M_i^2 (\mathbf{h h^\dagger})_{ii};  \\
\end{split}
\end{equation}
where $T_{\nu} \equiv  \tr \left[ \mathbf{h^\dagger h}\right] $.

The complete set of RGEs is given in the Appendix \ref{appendix: RGEs for Inert Doublet}. The last term of the RGE for the mass parameter for scalar $\eta$ namely $\mu_2^2$ shows its dependency on the RH neutrino mass term. For a larger value, this becomes the dominating term and pulls down the mass parameter, ultimately making it negative at higher energy. This in turns breaks the precious $\mathbb{Z}_2$ symmetry spoiling the model completely. If $M_i^2$ is of the same order as $\mu_2^2$, this outcome will not be realized, and $\mathbb{Z}_2$ will remain unbroken even at higher energies.
\subsection{Solution to the RGEs}
\begin{figure}
\centering
\begin{subfigure}{.5\textwidth}
  \centering
  \includegraphics[width=8cm]{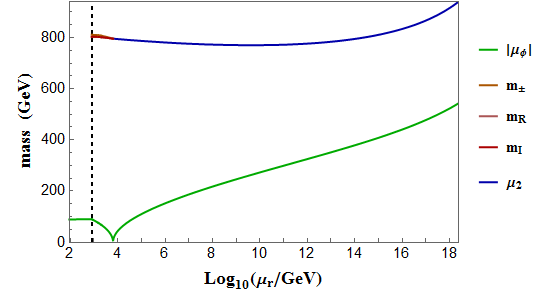}
  \caption{Evolution of mass parameters}
  \label{fig:sub1idm}
\end{subfigure}%
\begin{subfigure}{.45\textwidth}
  \centering
  \includegraphics[width=8.5cm]{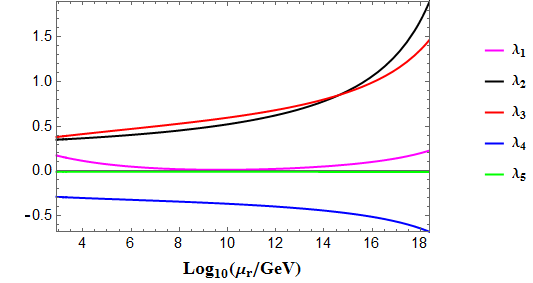}
  \caption{Evolution of quartic couplings}
  \label{fig:sub2idm}
\end{subfigure}
\caption{One-loop running of the parameters of scotogenic version of the inert doublet model from Planck scale down to weak scale. The black dashed line corresponds to the scale, $\mu_r=\mu_2$. In Fig. \ref{fig:sub1idm} the evolution of the absolute value of the SM Higgs mass parameter $\left(\left| \mu_\phi \right| \right)$ along with the masses of the components of the inert doublet has been plotted. The point at which $\left| \mu_\phi \right|$ touches the horizontal line corresponding to mass =$\;0 \; \text{GeV}$, is the energy scale where radiative EWSB is triggered as the sign of the SM Higgs mass-squared parameter $( \mu_\phi^2)$ switches from positive to negative while evolving from high to low energies. Note that $\mu_\phi^2$ turns negative around $\mu_r \approx  6\; \text{TeV}$, while $\mu_2^2$ remains positive all the way up to Planck scale emphasizing the fact that the $\mathbb{Z}_2$ remains unbroken. As below $\mu_r \approx 3 \; \text{TeV}$, electroweak symmetry has been broken, the common mass parameter $(\mu_2)$ for the components of inert doublet splits into $m_\pm$, $m_R$ and $m_I$.  In Fig. \ref{fig:sub2idm} shows the evolution of all the quartic couplings of the model from Planck scale down to weak scale. Note that the model remains perturbative all the way for the selected sample point.}
\label{fig:sample RGE inert doublet model}
\end{figure}

A sample point (given in the Table \ref{table: sample point inert doublet model}) generates the running of the mass parameters and the scalar quartic couplings shown in Fig. \ref{fig:sample RGE inert doublet model}. The sample point maintains all the boundedness conditions at all energy scales. The decoupling of the three RH nautrino was only considered for the running of the mass parameter $\mu_2^2$. As for all the other cases as the dependence on the RH neutrino mass is indirect, the decoupling effect is negligible.

\begin{table}[!ht]
\centering
\begin{center}\renewcommand\arraystretch{1.3}
\begin{tabular}{||c|c||c|c||} 
 \hline 
  \hline 
Quartic couplings & values &  Mass parameters & values\\ 
 \hline 
  \hline 
$\lambda_1(m_Z)$ & $0.258$&$\mu_2^2(\mu_2)$& $800^2\;\text{(GeV)}^2$\\
$\lambda_1(\mu_2)$ & $0.173$&$M_1$ & $900$ GeV\\
$\lambda_2(\mu_2)$ & $0.35$ & $M_2$ & $1500$ GeV\\
$\lambda_3(\mu_2)$ & $0.38$ &$M_3$ & $2000$ GeV\\
$\lambda_4(\mu_2)$&$-0.29$&$v(m_Z)$&$174.10$ GeV\\
$\lambda_5(\mu_2)$&$-0.01$  &$\mu_\phi^2(125 \;\text{GeV})$&$-(88.91)^2\;\text{(GeV)}^2$\\
    &&$\mu_\phi^2(\mu_2)$&$-(89.77)^2\;\text{(GeV)}^2$\\
\hline
  \hline 
 \end{tabular}  
 \end{center}
\caption{Quartic coupling and mass parameter values for the sample point used for the inert doublet model in Fig. \ref{fig:sample RGE inert doublet model}}
\label{table: sample point inert doublet model}
\end{table}
In Fig. \ref{fig:sample RGE inert doublet model}(a), below the energy level corresponding to the point where $\mu_\phi^2=0$ the electroweak symmetry is broken and the masses of the components of the scalar doublet $\eta$ are split. For the energies below that point the running of the masses of the charged and neutral components of the scalar $\eta$ is shown. Note that as the split in the masses for the sample point is very small $(\approx 6\; \text{GeV})$.  All the quartic couplings remain in the pertubative range and the new Yukawa couplings are chosen to be small $\mathbf{h}_{ij}\lesssim \mathcal{O}(.1)$.

\section{Scalar singlet dark matter model}
\label{sec:scalar singlet DM model}
Perhaps the simplest extension of SM requires the existence of a new heavy real scalar singlet of SM gauge group. An unbroken $\mathbb{Z}_2$ symmetry is assumed under which the singlet scalar is odd and can serve as a candidate for dark mater \cite{ssing}. Dark matter annihilation occurs efficiently in this model via Higgs portal interactions.
\subsection{The model}
In this simple extension of the SM, the added singlet can be protected from decaying into SM particles by virtue of a $\mathbb{Z}_2$ parity symmetry. This scenario can be well motivated from some higher symmetry at GUT scale where all the other additional particles lie above some intermediate  scale. For example, such a stable dark matter can be easily incorporated in $SO(10)$ models \cite{DM02}. In such cases, the low scale scalar potential becomes:
\begin{equation}
V(\phi, s)=\mu_\phi^2 \phi^\dagger \phi+\dfrac{\mu_s^2}{2}s^2+\dfrac{\lambda_1}{2}(\phi^\dagger \phi)^2+\dfrac{\lambda_2}{8}s^4+\dfrac{\lambda_3}{2}(\phi^\dagger \phi)s^2.
\end{equation}

Below the energy scale corresponding to the mass of the singlet, the effective quartic coupling is given by:
\begin{equation}
\label{Eq: Matching Condition singlet model}
 \lambda_1^{\text{eff}}=\lambda_1-\dfrac{\lambda_3^2}{\lambda_2}.
\end{equation}
And the mass of the observed Higgs particle is $m_h^2=2 \lambda_1^{\text{eff}} v^2$ and the matching condition Eq.(\ref{Eq: Matching Condition singlet model}) is needed while one evolves the RGE for the Higgs quartic coupling.
\subsection{The stability conditions and the evolution of the mass parameters}
The parameters of the scalar potential must obey the boundedness constraints so that the potential remains bounded from below. The conditions for this simple potential are given as:
\begin{equation}
\lambda_1\geq 0; \hspace{1cm} \lambda_2\geq 0; \hspace{1cm} \lambda_3\geq -\sqrt{\lambda_1 \lambda_2}.
\end{equation}
For this extension of SM, most of the RGEs of the SM remain the same. But one should update the RGEs for the Higgs quratic coupling $(\lambda_1)$  and the Higgs mass parameters $(\mu_\phi^2)$ along with the newly introduced quartic couplings $(\lambda_2, \lambda_3)$ and mass parameter $(\mu_s^2)$. The full set of new and updated RGEs is given in the Appendix \ref{appendix: RGEs for Scalar Extention}. The RGEs of the mass parameters are given as
\begin{equation}
\label{mass RGEs SM +real singlet scalar}
\begin{split}
16 \pi^2 \dfrac{d \mu_\phi^2}{d t} = & \; \[6 \lambda_1 +2 T -\dfrac{9}{10}g_1^2-\dfrac{9}{2}g_2^2\] \mu_\phi^2+\lambda_3 \mu_s^2; \\
16 \pi^2 \dfrac{d \mu_s^2}{d t}=& \; 3 \lambda_2 \mu_s^2+4 \lambda_3 \mu_\phi^2.\\
\end{split}
\end{equation}
From the RGEs of the mass parameters, one immediately notices that the coupling $\lambda_3$ has the potential to turn the mass parameter of SM Higgs negative at low energy while it remains positive at high energy. And one also notices that one needs a lower bound on coupling $\lambda_3$ to perform such a mechanism. The quartic coupling $\lambda_3$ is also the coupling that keeps the dark matter in thermal equilibrium. So a lower limit needed for the radiative electroweak symmetry breaking can be translated into a lower limit on the dark matter mass if one assumes that the thermal relic abundance of the dark matter is in agreement with the observed density, $\Omega_{DM}h^2 \simeq 0.1186$. Here $\Omega_{DM}$ is the critical mass density for dark matter and $h$ is the Hubble constant in units of $100$ km.(s.Mpc). The mass of the dark matter candidate is given by $m_s^2=m_{DM}^2=\dfrac{\lambda_3}{2}v^2+\mu_s^2$ and also assuming standard thermal freeze-out, we get $m_{DM}\simeq 3.3\; \lambda_3\;$ TeV.

Furthermore, according to Eq. (\ref{Eq:RGEs for SM extension}) the contribution of the quartic couplings $\lambda_3$ to the SM Higgs quartic coupling is just perfect to make the electroweak vaccum stable all the way to the Planck scale.
\begin{table}[!ht]
\centering
\begin{center}\renewcommand\arraystretch{1.3}
\begin{tabular}{||c|c||c|c||} 
 \hline 
  \hline 
Quartic couplings & values &  Mass parameters & values\\ 
 \hline 
  \hline 
$\lambda_1^{\text{eff}}(m_Z)$&$0.258$& $\mu^2_s$ & $560^2\;\text{(GeV)}^2$\\
  $\lambda_1^{\text{eff}}(\mu_s)$&$0.1887$&$ \; v(m_Z)$ & $174.10$ GeV\\
  $\lambda_1(\mu_s)$ & $0.247$&$\mu_\phi^2(125\; \text{GeV})$&$-(88.91)^2\;\text{(GeV)}^2$\\
 $\lambda_2(m_s)$ & $0.5$ &$\mu_\phi^2(\mu_s)$&$-(89.63)^2\;\text{(GeV)}^2$\\ 
$\lambda_3(m_s)$&$0.17$&&\\
\hline
  \hline 
 \end{tabular}  
 \end{center}
\caption{Quartic coupling and mass parameter values for the sample point used for the extension of SM by a real scalar singlet in Fig. \ref{fig:sample RGE SM_+ real singlet}. Note that the mass parameter of the dark matter candidate $\mu_s^2$ corresponds to the value needed for the right amount of thermal relic abundance.}
\label{table: sample point SM+real singlet}
\end{table}

\begin{figure}
\centering
\begin{subfigure}{.5\textwidth}
  \centering
  \includegraphics[width=8cm]{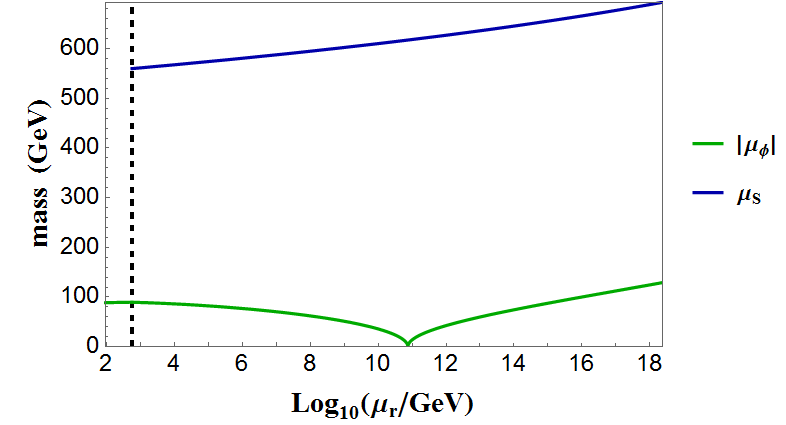}
  \caption{Evolution of mass parameters}
  \label{fig:sub1:sample RGE SM_+ real singlet}
\end{subfigure}%
\begin{subfigure}{.45\textwidth}
  \centering
  \includegraphics[width=8.5cm]{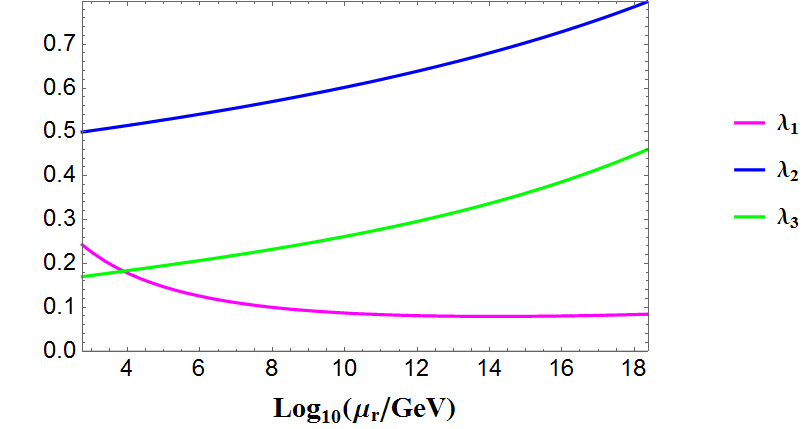}
  \caption{Evolution of quartic couplings}
  \label{fig:sub2:sample RGE SM_+ real singlet}
\end{subfigure}
\caption{One-loop running of the parameters of scalar singlet dark matter model from Planck scale down to weak scale. The black dashed line corresponds to the scale, $\mu_r=\mu_s$. In Fig. \ref{fig:sub1:sample RGE SM_+ real singlet} the evolution of the absolute value of the SM Higgs mass parameter $\left(\left| \mu_\phi \right| \right)$ along with the mass of the dark matter candidate has been plotted. The point at which $\left| \mu_\phi \right|$ touches the horizontal line corresponding to mass =$\;0 \; \text{GeV}$, is the energy scale where radiative EWSB is triggered as the sign of the SM Higgs mass-squared parameter $( \mu_\phi^2)$ switches from positive to negative while evolving from high to low energies. Note that $\mu_\phi^2$ turns negative around $\mu_r \approx 10^{11} \; \text{GeV}$, while $\mu_s^2$ remains positive all the way up to Planck scale emphasizing the fact that the $\mathbb{Z}_2$ that protects the dark matter candidate remains unbroken. In Fig. \ref{fig:sub2:sample RGE SM_+ real singlet} shows the evolution of all the quartic couplings of the model from Planck scale down to weak scale. Note that the model remains perturbative all the way for the selected sample point.}
\label{fig:sample RGE SM_+ real singlet}
\end{figure}
\begin{figure}[htb]
\begin{center}
\includegraphics[width=9cm]{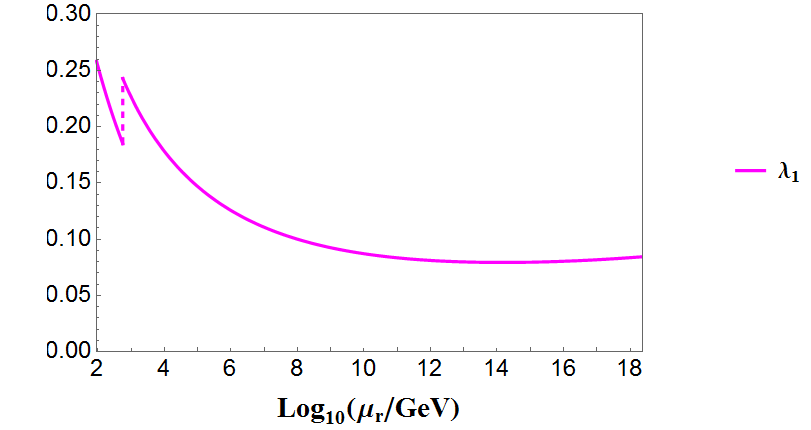}
\caption{Running of the SM Higgs quatic coupling in the extension of SM by a real scalar singlet. The discontinuous shift in the plot at the renormalization energy scale $\approx 560 \; \text{GeV}$ corresponds to the affect of the real scalar singlet. Below that energy scale, the effective Higgs quartic coupling given by Eq. (\ref{Eq: Matching Condition singlet model}) has been considered.}
\label{fig:Running of the SM Higgs quartic coupling}
\end{center}
\end{figure}
\subsection{Solution to the RGEs}
Like the previous cases, we evolved the SM couplings and parameters at two-loop level upto the energy scale corresponding to the mass of the singlet. From that point we evolved the new set of RGEs at one loop level upto Planck scale. 

We randomly took a sample point to illustrate the Radiative Electroweak symmetry breaking scenario for this extension of SM. The running in Fig. \ref{fig:sub1:sample RGE SM_+ real singlet} is from the mass of the singlet to the Planck scale, while Fig. \ref{fig:sub2:sample RGE SM_+ real singlet} is from weak scale to planck scale. To show the evolution of the SM Higgs quartic coupling we evolved the $\lambda_1^{\text{eff}}$ upto the singlet mass using two-loop SM RGEs and then used the matching condition in Eq. (\ref{Eq: Matching Condition singlet model}) and the set of updated RGEs to run the coupling upto Planck scale (see Fig. \ref{fig:Running of the SM Higgs quartic coupling}). This figure  also shows that the electroweak vacuum is perfectly stable for the selected sample point.


\section{Universal seesaw model with vector-like fermions}
\label{sec:Universal seesaw}
In universal seesaw models, one introduces a new set of heavy vector-like fermions which are responsible for the masses for quarks and charged leptons via a generalized seesaw mechanism \cite{Babu:1988mw}. Besides providing an explanation of the smallness of the masses of fermions like $u$, $d$, $e$, etc. in the context of left-right symmetry, $CP$- phenomenology study has revealed that such a model also harbors a solution for strong $CP$ problem as the $\overline{\theta}$ parameter of strong CP problem only has non-zero value $(\overline{\theta}\sim10^{-12})$ at two-loop level \cite{Babu:1988mw, Babu:1989rb, Mohapatra:2014qva, Dev:2015vjd}. Here we propose and analyze a $SM \times U(1)$ based universal seesaw model without replying on left-right symmetry. The strong CP problem may be solved via spontaneous CP violation \cite{Barrnelson}. The presence of new scalars needed for $U(1)$ symmetry breaking enables us to realize radiative EWSB.

\subsection{The model}
The model under scrutiny uses the seesaw mechanism for quarks and leptons and is based upon the assumption that there exists a set of TeV scale vector-like fermions. 
The original version of the model was constructed in the context of a left-right symmetric model \cite{Babu:1988mw}. Here we will study a variant of the model which is based on an extension of the SM gauge sector by an $U(1)_X$, where all the left-handed fermions of SM are neutral under the new gauge group while the TeV scale vector-like fermions are not. The model requires two more additional scalar bosons $(S_1\; \text{and}\;S_2)$ along with the SM Higgs doublet with the quantum charge assignment for all the particles of the model is listed in the Table \ref{table:particle content vector-like model}. While one singlet scalar $S_1$ or $S_2$ is sufficient for $U(1)$ symmetry breaking, both scalars are needed for seesaw mass generation.
\begin{table}[!ht]
\centering
\begin{center}\renewcommand\arraystretch{1.3}
\begin{tabular}{|c|c|c|}
 \hline 
Particle & $(SU(3)_C\times SU(2)_L\times U(1)_Y \times U(1)_X)$\\
\hline
$\mathcal{Q}$& $(3,2,\nicefrac{1}{6},0)$\\
$\mathcal{L}$& $(1,2,-\nicefrac{1}{2},0)$\\
$u^c$ & $(\overline{3},1,-\nicefrac{2}{3},-2)$\\
$d^c$ & $(\overline{3},1,\nicefrac{1}{3},2)$\\
$e^c$ & $(1,1,-1,2)$\\
$U$ & $(3,1,\nicefrac{2}{3},1)$\\
$U^c$ & $(\overline{3},1,-\nicefrac{2}{3},1)$\\
$D$ & $(3,1,-\nicefrac{1}{3},-1)$\\
$D^c$ & $(\overline{3},1,\nicefrac{1}{3},-1)$\\
$E$ & $(1,1,1,-1)$\\
$E^c$ & $(1,1,-1,-1)$\\
$\phi$ & $(1,2,\nicefrac{1}{2},-1)$\\
$S_1$ & $(1,1,0,1)$\\
$S_2$ & $(1,1,0,-2)$\\
 \hline 
 \end{tabular}  
 \caption{Particle content of the vector-like fermion model}
 \label{table:particle content vector-like model}
 \end{center}
\end{table}

The VEV of the singlet $S_2$ gives masses to the vector-like fermions and the VEV of $S_1$ along with the electroweak VEV mixes the right and left-handed quarks (and leptons) with the vector-like quarks (and leptons) while the $U(1)_X$ symmetry forbids the bare mass terms of any of the vector-like fermions. Thus, this model is natural framework for universal seesaw mechanism related without left-right symmetry.

Note that the setup is anomaly free. As the added fermions are vector-like, most of the anomalies cancel trivially. The only non-trivial cancellations are for the cases : $U(1)_Y \left[ U(1)_X \right]^2$, $\left[U(1)_Y\right]^2  U(1)_X$ and $\tr \left[ U(1)_X \right]$. A straightforward calculation using Table \ref{table:particle content vector-like model} shows that the anomalies for these three cases are all zero.

The Yukawa sector of the Lagrangian for this model is given by:
\begin{equation}
\begin{split}
\mathcal{L}_Y =& \;\mathbf{Y_u} \mathcal{Q}U^c \phi+ \mathbf{F_u} U u^c S_1+\mathbf{G_u} U U^c S_2 \\ & -\mathbf{Y_d} \mathcal{Q}D^c \tilde{\phi}+\mathbf{F_d} D d^c S_1^*+\mathbf{G_d} D D^c S_2^*\\ & -\mathbf{Y_e}\mathcal{L}E^c \tilde{\phi}+\mathbf{F_e} E e^c S_1^*+\mathbf{G_e} E E^c S_2^*+h.c.
\end{split}
\end{equation}
where
\begin{equation}
\begin{split}
\mathbf{Y_u} \mathcal{Q}U^c \phi =& (\mathbf{Y_u})_{ij}(u_i U^c_j \phi^0-d_iU_j^c\phi^+_u);\\
-\mathbf{Y_d} \mathcal{Q}D^c \tilde{\phi} =& (\mathbf{Y_d})_{ij}(u_i D^c_j \phi^- +d_iD_j^c \overline{\phi}^0);\\
-\mathbf{Y_e} \mathcal{L}E^c \tilde{\phi} =& (\mathbf{Y_e})_{ij}(\nu_i E^c_j \phi^- +e_iE_j^c \overline{\phi}^0);\\
\end{split}
\end{equation}
and \begin{equation}
\phi=\begin{pmatrix}
\phi^+\\
\phi^0
\end{pmatrix};\hspace{1cm}\tilde{\phi}=\begin{pmatrix}
\overline{\phi}^0\\
\phi^-
\end{pmatrix}
\end{equation}

When the electroweak doublet and scalar singlets both get VEVs, one acquires the fermion mass matrix $\mathcal{M}_f$ in the seesaw form for both quark and lepton  as,
\begin{equation}
\mathcal{M}_f=\begin{pmatrix}
0 & \dfrac{1}{\sqrt{2}}\mathbf{Y_f}v\\
\mathbf{F_f}v_{s_1} & \mathbf{G_f} v_{s_2}\\
\end{pmatrix}
\end{equation}
where $\mathbf{f}=\mathbf{u,d,e}$. For such a case the mass of the light quark (or lepton) becomes $m_f \approx \dfrac{Y_f F_fv v_{s_1}}{\sqrt{2}G_fv_{s_2}}$. Since these masses scale quadratically with Yukawa couplings, fermion mass hierarchy may be explained with only a mild hierarchy $\sim (10^{-2}- 10^{-3})$ in the Yukawa couplings.

The scalar potential of the model can be written as
\begin{equation}
\begin{split}
V(\phi,S_1,S_2)= &\;\mu_\phi^2 \phi^\dagger \phi +\mu_1^2 S_1^*S_1+\mu_2^2 S_2^*S_2-\l( \mu S_1^2 S_2+h.c. \r) +\dfrac{\lambda_1}{2}(\phi^\dagger \phi)^2+\dfrac{\lambda_2}{2} (S_1^*S_1)^2\\&+\dfrac{\lambda_3}{2} (S_2^*S_2)^2+\lambda_4 (\phi^\dagger \phi)(S_1^* S_1)+\lambda_5 (\phi^\dagger \phi)(S_2^* S_2)+\lambda_6 (S_1^* S_1)(S_2^* S_2).
\end{split}
\end{equation}
Here $\mu$ can be taken as real without any loss of generality by the redefinition of the complex scalar $S_2$ .

When the scalar $S_1$ develops a VEV $v_{s_1}$ via radiative corrections (see below), the $S_2$ develops an induced VEV due to the linear term in $S_2$ in the potential. For such a case, the imaginary part of the complex scalar $S_1$ is absorbed by the broken generator of $U(1)_X$ and the mass matrix for the scalar becomes:
\begin{equation}
\mathcal{M}^2_s=\begin{pmatrix}
2 \lambda_1 v^2 & 2 \lambda_4 v v_{s_1} & 0 & 0\\
2 \lambda_4 v v_{s_1} & 2 \lambda_2 v_{s_1}^2 & - 2 \mu v_{s_1}& 0 \\
0 & - 2 \mu v_{s_1}&\lambda_5 v^2 +\lambda_6 v_{s_1}^2 +\mu_2^2 &0 \\
0&0&0&\lambda_5 v^2 +\lambda_6 v_{s_1}^2 +\mu_2^2
\end{pmatrix}
\end{equation}
Here the basis of the matrix $\mathcal{M}^2_s$ is $\{ m_h,m_{S_1},m_{S_{2R}},m_{S_{2I}}\}$, where $m_h$ is the SM Higgs, and $m_{S_1}$ is the mass of the singlet $S_1$ and the $m_{S_{2R}},m_{S_{2I}}$ are the masses of the real and imaginary part of the $S_2$ scalar. From the potential we find that the induced VEV for the scalar $S_2$ is given by:
\begin{equation}
v_{s_2}=\dfrac{\sqrt{2}\mu v^2_{s1}}{\lambda_5 v^2 +\lambda_6 v_{s_1}^2+\mu_2^2}.
\end{equation}
Small value of the coupling $\lambda_4$ and $v_{s_1}$ around TeV scale will ensure a small mixing between the SM Higgs and the singlet $S_1$ while the mixing between $S_1$ and $S_2$ depends on the cubic coupling parameter $\mu$. 
\subsection{The stability condition and the evolution of the mass parameters}
From the stability point of view, the scalar potential of the vector-like fermion model and the two-loop neutrino mass model are identical. So, the stability condition given by Eq. (\ref{eq:stability conditions two-loop}) is applicable here too.

The RGEs for the mass parameters are found to be
\begin{equation}
\label{mass RGEs for vector-like Fermion model}
\begin{split}
16 \pi^2 \dfrac{d \mu_\phi^2}{d t} =&\; \mu_\phi^2 \[ 6 \lambda_1 - \dfrac{9}{10}g_1^2-\dfrac{9}{2}g_2^2-6 g_4^2+2 T\]+2 \lambda_4 \mu_1^2+2 \lambda_5 \mu_2^2;\\
16 \pi^2 \dfrac{d \mu_1^2}{d t} =&\; \mu_1^2 \[ 4 \lambda_2 - 6g_4^2+2 T_F\]+4 \lambda_4 \mu_\phi^2+2 \lambda_6 \mu_2^2+8 \mu^2;\\
16 \pi^2 \dfrac{d \mu_2^2}{d t} =&\; \mu_2^2 \[4 \lambda_3 - 24g_4^2+2 T_G\]+4 \lambda_5 \mu_\phi^2+4 \lambda_6 \mu_1^2+4 \mu^2;\\
16 \pi^2 \dfrac{d \mu}{d t} =&\; \mu \[ 2 \lambda_2+2 \lambda_6 - 18g_4^2+2 T_F+T_G\].\\
\end{split}
\end{equation}
The complete set of RGEs are given in the Appendix \ref{appendix: RGEs for vector-like Fermion}.
\begin{table}[!ht]
\centering
\begin{center}\renewcommand\arraystretch{1.3}
\begin{tabular}{||c|c||c|c||} 
 \hline 
  \hline 
 Quartic couplings & values &  mass parameters & values\\ 
 \hline 
  \hline 
  $\lambda_1(m_Z)$ & $0.258$&$\mu_1^2(\mu_s)$ & $-(1800)^2\;\text{(GeV)}^2$ \\
  $\lambda_1(\mu_s)$ & $0.175$&$\mu_2^2(\mu_s)$ & $(2550)^2\;\text{(GeV)}^2$ \\
  $\lambda_2(\mu_s)$&$0.25$&$ \mu_\phi^2(\mu_s)$ & $-(89.74)^2\;\text{(GeV)}^2$\\
  $\lambda_3(\mu_s)$&$0.24$&$\mu(\mu_s)$ & $850\;\text{GeV}$\\
  $\lambda_4(\mu_s)$&$0.02$ &$\mu_\phi^2(125 \; \text{GeV})$&$-(88.91)^2\;\text{(GeV)}^2$\\
  $\lambda_5(\mu_s)$&$0.1$&$(G_u)_{ii}(M_z)\sim(G_d)_{ii}(M_z)\sim(G_e)_{ii}(M_z)$&$0.45$\\
  $\lambda_6(\mu_s)$&$0.09$&$v_{s_1}$&$3.60$ TeV\\
  $\mu_s$&$750\;\text{GeV}$&$v_{s_2}$&$2.26$ TeV\\
\hline
  \hline 
 \end{tabular}  
 \end{center}
\caption{Quartic couplings and mass parameter values for the sample point used for the vector-like fermion model in Fig. \ref{fig:sample RGE vector-like model}}
\label{table: sample point vector-like fermion model}
\end{table}

\begin{figure}
\centering
\begin{subfigure}{.5\textwidth}
  \centering
  \includegraphics[width=8cm]{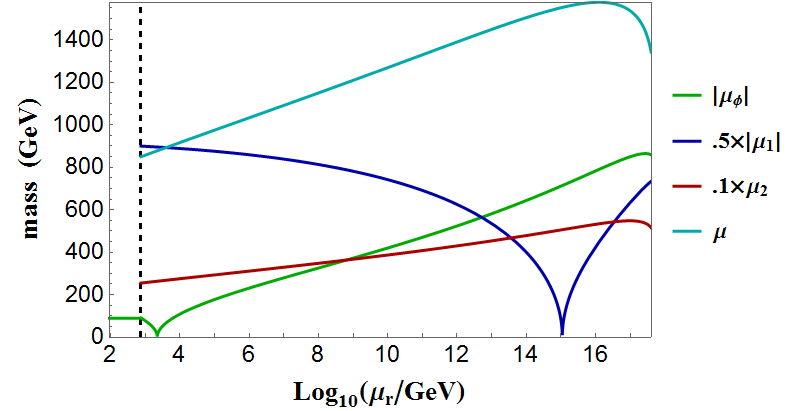}
  \caption{Evolution of mass parameters}
  \label{fig:sub1VF}
\end{subfigure}%
\begin{subfigure}{.5\textwidth}
  \centering
  \includegraphics[width=8.5cm]{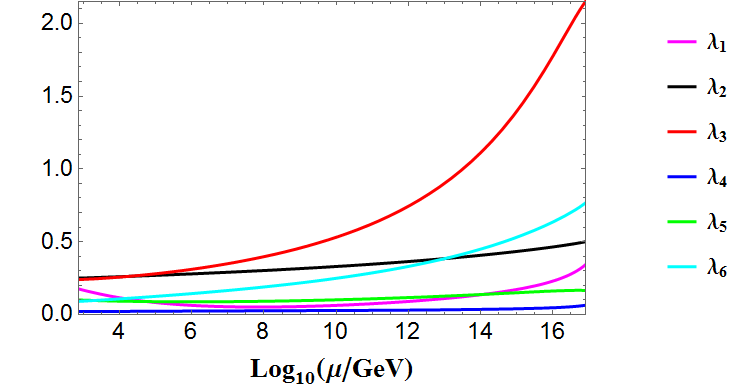}
  \caption{Evolution of quartic couplings}
  \label{fig:sub2VF}
\end{subfigure}
  \caption{One-loop running of the couplings and mass parameters of vector-like fermion model. The black dashed line corresponds to the scale, $\mu_r=\mu_s$. In Fig. \ref{fig:sub1VF} the evolution of the absolute value of the SM Higgs mass parameter $\left(\left| \mu_\phi \right| \right)$ and $\left| \mu_1 \right|$ along with the masses parameters $\mu_2$ and $\mu$ have been plotted. The point at which $\left| \mu_\phi \right|$ touches the horizontal line corresponding to mass =$\;0 \; \text{GeV}$, is the energy scale where radiative EWSB is triggered as the sign of the SM Higgs mass-squared parameter $( \mu_1^2)$ switches from positive to negative while evolving from high to low energies. For the sample point, this happens around TeV energy scale. Note that the mass parameter of the scalar $S_1$ also turns negative around the renormalization energy scale $\approx 10^{15}\; \text{GeV}$ indicating that the $U(1)_X$ symmetry also gets broken radiatively. In Fig. \ref{fig:sub2VF} shows the evolution of all the quartic couplings of the model from Planck scale down to weak scale. Note that the model remains perturbative all the way for the selected sample point.}
\label{fig:sample RGE vector-like model}
\end{figure}
\subsection{Solution to the RGEs}
To find the solution of the set of RGEs, we took a more simplified case where we kept all the Yukawa coupling $\mathbf{F}$ to be small and negligible and Yukawa coupling $\mathbf{G}\simeq \mathcal{O}(1)$. The numerical solution was hunted for the case where one of the eigenvalue of the scalar mass matrix $\mathcal{M}_s$ corresponds to the SM Higgs boson and another one corresponds to scalar boson of mass $\sim \mu_s$. Here, we used $\mu_s=750$ GeV just as an example as similar universal seesaw model \cite{Dev:2015vjd} was used to explain apparent diphoton excess \cite{750found} which eventually became statistically insignificant \cite{750gone}. The vector-like fermion mass was kept around TeV scale where the mass is approximated by $\sim \mathbf{G}v_{s_2}$. One such sample point is given by Table \ref{table: sample point vector-like fermion model}. As the first new particle in this model is at $\mu_s=750$ GeV, the SM RGEs were evolved at two-loop level upto the scale $\mu_s$ and then the new set of RGEs was deployed to do the evolution of the couplings and mass parameters. Fig. \ref{fig:sample RGE vector-like model} shows that the both the vevs (electroweak VEV and VEV for the single $S_1$) can be generated by radiative correction.

\section{Conclusion}
\label{sec:conclusion}
We have presented in this paper various extensions of the SM where electroweak symmetry breaking is triggered by the renormalization group flow. Even though such symmetry breaking fails to occur in SM, the scalar extensions are able to incorporate this attractive mechanism. Extensions like Type-II seesaw models, loop induced neutrino mass models and scalar dark matter models all have this built-in feature. A common shared feature of all models where radiative EWSB is realized is the presence of new scalars at the TeV scale. These TeV scale scalars may be detected in the Large Hadron Collider (LHC) in the near future.
\section*{Acknowledgments}
KSB would like to thank Bartol Research Institute for hospitality during a visit where the work was initiated. KSB and SK would like to thank the organizers of CETUP* 2016 for hospitality and the  participants for helpful discussions and comments. The work of KSB and SK is supported in part by the U. S. Department of Energy Grant No.  de-sc0016013 and the work of IG is supported in part by  Bartol Research Institute. 
\begin{appendices}
\appendixpage
\section{Boundedness condition for type-II seesaw}
\label{appendix: stability cond}
The gauge transformation of the fields defined in Eqs. (\ref{eq:type-II scalar potential})-(\ref{eq:type II Yukawa}) can be written as:
\begin{equation}
\ell \rightarrow \mathcal{U} \ell\; ; \hspace{3cm} \Delta \rightarrow \mathcal{U} \Delta \mathcal{U}^\dagger
\end{equation}
where $\mathcal{U}$ is a unitary matrix. Under this gauge transformation the Yukawa term remains invariant:
\begin{equation}
\mathcal{L}_Y\supset \ell^T i \sigma_2 \Delta \ell \rightarrow \ell^T \mathcal{U}^T i \sigma_2 \mathcal{U} \Delta \mathcal{U}^\dagger \mathcal{U}\ell = \ell^T i \sigma_2 \Delta \ell.
\end{equation}

Let us define $\hat{\Delta}=i \sigma_2 \Delta$:
\begin{equation}
\hat{\Delta}=\begin{pmatrix}
\Delta^0 & -\dfrac{\Delta^+}{\sqrt{2}}\\
-\dfrac{\Delta^+}{\sqrt{2}} & - \Delta^{++}
\end{pmatrix}.
\end{equation}
The gauge transformation of the field $\hat{\Delta}$ can be written as
\begin{equation}
\hat{\Delta}\rightarrow i \sigma_2 \mathcal{U}\Delta \mathcal{U}^\dagger=\mathcal{U}^*\hat{\Delta}\mathcal{U}^\dagger
\end{equation}
Now, one can diagonalize $\hat{\Delta}$ by gauge transformation and as $\tr \left(\mathcal{U}^* \hat{\Delta} \mathcal{U}^\dagger\right) \neq 0$, we can write $\hat{\Delta}$ in such a basis as
\begin{equation}
\hat{\Delta}= \begin{pmatrix}
a & 0 \\
0 & b \; e^{i \alpha}
\end{pmatrix} \; ; \hspace{3cm} \text{where} \; a,b \; \text{and}\;\alpha \; \text{are real}.
\end{equation}

Let us define the Higgs doublet as
\begin{equation}
\phi=\begin{pmatrix}
c \; e^{i \delta}\\
d \; e^{i \gamma}
\end{pmatrix} \; ; \hspace{3cm} \text{where} \; c,d,\delta \; \text{and}\;\gamma \; \text{are real}.
\end{equation}
In terms of these real fields the quartic part of the  scalar potential (given in Eq. (\ref{eq:type-II scalar potential}))becomes
\begin{equation}
V^{(4)}=\dfrac{\lambda_1}{2}\; u^4+\dfrac{\lambda_2}{2}\;\left( a^2 +b^2 \right)^2+\lambda_3 \; a^2 b^2+\lambda_4 \; u^2 \left( a^2+b^2 \right) +\lambda_5 \; \left( a^2 -b^2 \right) \cos 2\beta
\end{equation}
where
\begin{equation}
c=u \cos\beta \; ; \hspace{3cm} d=u \sin\beta.
\end{equation}
The quartic couplings of the potential form a vector space spanned by the real valued vector, $\bm{x}^T=\left( u^2, \; a^2,\; b^2\right)$ and the quartic couplings of the scalar potential can be written as
\begin{equation}
V=\dfrac{1}{2}\;  \boldmath{x}^T \boldmath{\lambda\, x},
\end{equation}
where
\begin{equation}
\bm{\lambda}=\begin{pmatrix}
\lambda_1 & \lambda_4+\lambda_5 \cos 2\beta & \lambda_4-\lambda_5 \cos 2\beta \\
\lambda_4+\lambda_5 \cos 2\beta & \lambda_2 &\lambda_2+\lambda_3 \\
\lambda_4-\lambda_5 \cos 2\beta & \lambda_2+\lambda_3 & \lambda_2
\end{pmatrix}.
\end{equation}
We use known results from the copositivity conditions of real symmetric matrices \cite{Hadeler:1983, Klimenko:1985} to determine the boundedness conditions as
\begin{enumerate}[label=(\roman*)]
\item \hfill  \makebox[5pt][r]{%
            \begin{minipage}[b]{\textwidth}
              \begin{equation}
                 \lambda_1 \geq 0 \; ; \hspace{.2cm} \lambda_2 \geq 0\; ; \hspace{10cm}
              \end{equation}
          \end{minipage}}
 \item \hfill  \makebox[5pt][r]{%
            \begin{minipage}[b]{\textwidth}
              \begin{equation}
                 \lambda_4+\lambda_5 \cos 2 \beta  \geq -\sqrt{\lambda_1 \lambda_2} \; ; \hspace{.2cm}  \lambda_4-\lambda_5 \cos 2 \beta  \geq -\sqrt{\lambda_1 \lambda_2} \; ; \hspace{.2cm} \lambda_2+\lambda_3 \geq -\lambda_2; \hspace{.7cm}
              \end{equation}
          \end{minipage}}
 \item \hfill  \makebox[5pt][r]{%
            \begin{minipage}[b]{\textwidth}
              \begin{equation}
                 2 \lambda_4 \sqrt{\lambda_2}+2 \lambda_2 \sqrt{\lambda_1} +\lambda_3 \; \sqrt{\lambda_1} \geq 0 \;  \hspace{.2cm} \text{or} \det \bm{\lambda} \geq 0\; ; \hspace{4.5cm}
              \end{equation}
          \end{minipage}}
\end{enumerate}
where 
\begin{equation}
\det \bm{\lambda}=-2 \lambda_1 \lambda_2 \lambda_3 -\lambda_1 \lambda_3^2+2 \lambda_3 \lambda_4^2-2\left(2 \lambda_2 +\lambda_3\right) \lambda_5^2 \cos^2 2\beta.
\end{equation}
As this set of boundedness conditions need to be satisfied for all values of $\beta$, this set of conditions reduces to the set of inequalities (\ref{cond1}, \ref{cond2} and \ref{cond3}). Note that, all the conditions mentioned in inequalities (\ref{cond1}) and (\ref{cond2}) and at least one of the conditions in inequality (\ref{cond3}) need to be satisfied for the potential to be bounded from below.
\section{Complete set of RGEs for Type-II neutrino mass model}
\label{appendix: RGEs for Type-II}
The RGEs for the Yukawa couplings are given by \cite{Schmidt:2007nq}:
\begin{equation}
\label{RGEs Type-II}
\begin{split}
 16\pi^2 \dfrac{d\mathbf{Y}_d}{d t}  = &  \mathbf{Y_d} \left[ \frac{3}{2} \mathbf{Y_d^\dagger Y_d} -\frac{3}{2}\, \mathbf{Y_u^\dagger Y_u}\right]+ \mathbf{Y_d}\left[ T- \frac{1}{4} g_1^2 - \frac{9}{4} g_2^2 - 8\,g_3^2\right]; \\
16\pi^2 \dfrac{d\mathbf{Y}_u}{d t}  = & \mathbf{Y_u} \left[ \frac{3}{2} \mathbf{Y_u^\dagger Y_u} - \frac{3}{2}\, \mathbf{Y_d^\dagger Y_d}\right]+\mathbf{Y_u}\left[ T-\frac{17}{20} g_1^2 - \frac{9}{4} g_2^2 - 8\,g_3^2\right]; \\
16\pi^2 \dfrac{d \mathbf{Y}_e}{d t}=&\mathbf{Y_e}\left[\frac{3}{2}\mathbf{Y_e^\dagger Y_e}+\frac{3}{2}\mathbf{Y_\Delta^\dagger Y_\Delta}\right] +\mathbf{Y_e}\left[T-\frac{9}{4}g_1^2-\frac{9}{4}g_2^2\right];\\
16\pi^2 \dfrac{d \mathbf{Y}_\Delta}{d t}=&\left[\frac{1}{2}\mathbf{Y_e^\dagger Y_e}+\frac{3}{2}\mathbf{Y_\Delta^\dagger Y_\Delta}\right]^T \mathbf{Y_\Delta} +\mathbf{Y_\Delta}\left[\frac{1}{2}\mathbf{Y_e^\dagger  Y_e}+\frac{3}{2}\mathbf{Y_\Delta^\dagger Y_\Delta}\right] \\&+\left[-\frac{3}{2}\left(\frac{3}{5}g_1^2+3g_2^2\right)+\tr\left(\mathbf{Y_\Delta^\dagger  Y_\Delta}\right)\right]\mathbf{Y_\Delta}.\\
\end{split}
\end{equation}
The RGEs for the quartic couplings of the Lagrangians are given by \cite{Schmidt:2007nq}:
\begin{equation}
\begin{split}
16\pi^2\dfrac{d \lambda_1}{d t} = &12 \lambda_1^2
  -3\lambda_1\left(3g_2^2+\frac{3}{5}g_1^2\right)+3g_2^4+\frac{3}{2}\left(\frac{3}{5}g_1^2+g_2^2\right)^2+4\lambda_1 T -8 H +12\lambda_4^2+8\lambda_5^2;\\
16\pi^2 \dfrac{d \lambda_2}{d t}=&-\frac{36}{5}g_1^2\lambda_2-24g_2^2\lambda_2+\frac{108}{25}g_1^4 +18g_2^4+\frac{72}{5}g_1^2g_2^2 +14\lambda_2^2 +4\lambda_2\lambda_3+2\lambda_3^2+4\lambda_4^2+4\lambda_5^2\nonumber\\
&+4\tr\left(\mathbf{Y_\Delta^\dagger Y_\Delta}\right)\lambda_2 -8\tr\left(\mathbf{Y_\Delta^\dagger Y_\Delta Y_\Delta^\dagger Y_\Delta}\right);\\
16\pi^2\dfrac{d\lambda_3}{d t}=&-\frac{36}{5}g_1^2\lambda_3-24g_2^2\lambda_3 +12g_2^4-\frac{144}{5}g_1^2g_2^2+3\lambda_3^2+12\lambda_2\lambda_3-8\lambda_5^2 +4\tr\left(\mathbf{Y_\Delta^\dagger Y_\Delta}\right)\lambda_3\nonumber\\
&+8\tr\left(\mathbf{Y_\Delta^\dagger Y_\Delta Y_\Delta^\dagger Y_\Delta}\right);\\
16\pi^2 \dfrac{d \lambda_4}{d t}=&-\frac{9}{2}g_1^2\lambda_4-\frac{33}{2}g_2^2\lambda_4+\frac{27}{25}g_1^4+6g_2^4+\[8\lambda_2+2\lambda_3+6\lambda_1+4\lambda_4+2 T + 2\tr\left(\mathbf{Y_\Delta^\dagger Y_\Delta}\right)\]\lambda_4\nonumber +8\lambda_5^2;\\
16\pi^2 \dfrac{d \lambda_5}{d t}=&-\frac{9}{2}g_1^2\lambda_5-\frac{33}{2}g_2^2\lambda_5-\frac{18}{5}g_1^2g_2^2+\[2\lambda_2-2\lambda_3+2\lambda_1+8\lambda_4+2 T + 2 \tr\left(\mathbf{Y_\Delta^\dagger Y_\Delta}\right)\]\lambda_5.\nonumber\\
\end{split}
\end{equation}
\begin{equation}
\begin{split}
\text{where}\hspace{1cm} T &=\tr \[\mathbf{Y_e^\dagger Y_e}  +3 \mathbf{Y_d^\dagger Y_d} +3 \mathbf{Y_u^\dagger Y_u} \] ;\\
H &= \tr \[\mathbf{Y_e^\dagger Y_e Y_e^\dagger Y_e}  +3 \mathbf{Y_d^\dagger Y_d Y_d^\dagger Y_d} +3 \mathbf{Y_u^\dagger Y_u Y_u^\dagger Y_u} \].
\end{split}
\end{equation} 
The RGEs for the mass parameters are given by Eq.( \ref{mass RGEs Type-II}).\\

Beyond the energy scale corresponding to the mass of the triplet $(\mu_\Delta)$ the SM gauge coupling evolution also needs to be recalculated due to the triplet's contribution. While the weak triplet does not effect the evolution of the $SU(3)_C$ gauge coupling evolution, it does change the RGEs of the other gauge couplings. The RGEs for the gauge couplings are given by:
\begin{equation}
\label{RGEs Type-II gauge}
16 \pi^2 \dfrac{d g_i}{d t}  = b_i \; g_i^3,
\end{equation}
where $g_i = \{ g_3,g_2,g_1\}$ are the three gauge couplings with the one loop $\beta$-function coefficient $b_i=\{-7,-\nicefrac{5}{2},\nicefrac{47}{10}\}$.

\section{Complete set of RGEs for two-loop neutrino mass model}
\label{appendix: RGEs for Two-loop}
For the Two-loop neutrino mass model, among the gauge couplings only the hypercharge gauge coupling is modified due the additional scalar particles. So, the RGEs for the SM gauge couplings are given by:
\begin{equation}
16 \pi^2 \dfrac{d g_i}{d t} =b_ig_i^3,
\end{equation}
where $g_i = \{ g_3,g_2,g_1\}$ are the three gauge couplings with the one loop $\beta$-function coefficient $b_i=\{-7,-\nicefrac{19}{6},\nicefrac{51}{10}\}$\\

The RGEs for the Yukawa couplings are given by \cite{Babu:2014kca}:
\begin{equation}
\begin{split}
16 \pi^2 \dfrac{d\mathbf{h}}{d t} &= 4(\mathbf{hh^\dagger h})+4 \mathbf{h}\tr(\mathbf{h^\dagger h})-\dfrac{18}{5} g_1^2\mathbf{h}+\dfrac{1}{2}(\mathbf{h Y^\dagger_\ell Y_\ell})+\dfrac{1}{2}(\mathbf{Y^T_\ell Y^*_\ell h});\\
16 \pi^2 \dfrac{d\mathbf{f}}{d t} &= 4(\mathbf{ff^\dagger f})+4 \mathbf{f}\tr(\mathbf{f^\dagger f})+\dfrac{1}{2}(\mathbf{f Y_\ell Y^\dagger_\ell })+\dfrac{1}{2}(\mathbf{ Y^*_\ell Y^T_\ell f})-\dfrac{3}{2} \mathbf{f}(-\dfrac{3}{5}g_1^2+g^2_2).\\
\end{split}
\end{equation}
The RGEs for the quartic scalar couplings are given by \cite{Babu:2014kca, Chao:2012xt}:
\begin{equation}
\begin{split}
16 \pi^2 \dfrac{d \lambda_1}{d t} =& \;12 \lambda_1^2+2 \lambda_4^2+2\lambda_5^2-\lambda_1\l(9g_2^2+\dfrac{9}{5} g_1^2\r)+\dfrac{9}{4} g_2^4+\dfrac{27}{100} g_1^4+\dfrac{9}{10} g_2^2 g_1^2+4 \lambda_1 T -4 H;\\
16 \pi^2 \dfrac{d\lambda_2}{d t} =& \;10\lambda_2^2+4 \lambda_4^2+2\lambda_6^2-\dfrac{36}{5} \lambda_2 g_1^2+\dfrac{108}{25} g_1^4 +16 \lambda_2 \tr (\mathbf{f^\dagger f}) - 32\; \tr(\mathbf{f^\dagger f})^2;\\
16 \pi^2 \dfrac{d\lambda_3}{d t}= & \;10\lambda_3^2+4 \lambda_5^2+2\lambda_6^2-\dfrac{144}{5} \lambda_2 g_1^2+\dfrac{864}{25} g_1^4 +16 \lambda_3 \tr (\mathbf {h^\dagger h}) - 64 \; \tr(\mathbf{h^\dagger h})^2;\\
16 \pi^2 \dfrac{d\lambda_4}{d t}=&\; 6 \lambda_1 \lambda_4+4 \lambda_2 \lambda_4+2 \lambda_5 \lambda_6 +4 \lambda_4^2-\lambda_4 \l( \dfrac{9}{2} g_2^2+\dfrac{45}{10} g_1^2 \r)+\dfrac{27}{50} g_1^4\\ & +2 \lambda_4 \[4 \tr ( \mathbf{f^\dagger f})+T\]-8 \tr \l(\mathbf{f^\dagger f Y^\dagger_\ell Y_\ell}\r); 
\end{split}
\end{equation}
\begin{equation}
\begin{split}
16 \pi^2\dfrac{d\lambda_5}{d t}=&\; 6 \lambda_1 \lambda_5+4 \lambda_3 \lambda_5+2 \lambda_4 \lambda_6 +4 \lambda_5^2-\lambda_5 \l( \dfrac{9}{2} g_2^2+\dfrac{153}{10} g_1^2 \r)+\dfrac{108}{25} g_1^4\\ & +2 \lambda_5 \[4 \tr ( \mathbf{h^\dagger h})+T\]-8 \tr \l(\mathbf{Y^\dagger_\ell Y_\ell h^\dagger h }\r); \\
16 \pi^2 \dfrac{d\lambda_6}{d t}=&\; 4 \lambda_2 \lambda_6+4 \lambda_3 \lambda_6+4 \lambda_4 \lambda_5 +4 \lambda_6^2-\dfrac{90}{5} \lambda_6 g_1^2 +\dfrac{432}{25} g_1^4 +8 \lambda_6 \[ \tr ( \mathbf{f^\dagger f})+\tr ( \mathbf{h^\dagger h})\].\\
\end{split}
\end{equation}
The RGEs for the mass parameters are given by Eq. (\ref{mass RGEs two-loop neutrino}).

\section{Complete set of RGEs for inert doublet model}
\label{appendix: RGEs for Inert Doublet}
The one loop RGEs for the Inert doublet model have already been computed. The SM gauge coupling RGEs are given by:
\begin{equation}
16 \pi^2 \dfrac{d g_i}{d t} = b_i g_i^3,
\end{equation}
where $b_i=(-7,-3,\dfrac{21}{5}) $ are the $\beta$-coefficients of the SM gauge couplings updated with the added particles.  \\

The quark sector of the model remains unchanged, while the leptonic sector needs to be revisited. The RGEs for the leptonic Yukawa couplings are \cite{Merle:2015gea}:
\begin{equation}
\begin{split}
16 \pi^2 \dfrac{d\mathbf{Y_e}}{d t}=&\mathbf{Y_e} \{\dfrac{3}{2}\mathbf{Y_e^\dagger Y_e} +\dfrac{1}{2}\mathbf{h^\dagger h}+T -\dfrac{9}{4}g_1^2-\dfrac{9}{4}g_2^2\};\\
16 \pi^2 \dfrac{d\mathbf{h}}{d t}=&\mathbf{h} \{\dfrac{3}{2}\mathbf{h^\dagger h} +\dfrac{1}{2}\mathbf{Y_e^\dagger Y_e}+T_\nu -\dfrac{9}{20}g_1^2-\dfrac{9}{4}g_2^2\};\\
16 \pi^2 \dfrac{d\mathbf{M}}{d t}=& \{ (\mathbf{h h^\dagger}) \mathbf{M} +\mathbf{M} (\mathbf{h h^\dagger})^* \}.
\end{split}
\end{equation}

For the quartic scalar coupling we find the following set of RGEs \cite{Hill:1985tg}:
\begin{equation}
\begin{split}
16 \pi^2 \dfrac{d \lambda_1}{d t} =& 12 \lambda_1^2+4 \lambda_3^2+4 \lambda_3 \lambda_4+2 \lambda_4^2+2 \lambda_5^2+\dfrac{3}{4}\l(\dfrac{9}{25} g_1^4+\dfrac{6}{5} g_1^2g_2^2+3g_2^4\r) \\ & - 3 \lambda_1\l(\dfrac{3}{5} g_1^2+3g_2^2\r) +4 \lambda_1 T -4 H;\\
16 \pi^2 \dfrac{d \lambda_2}{d t} =& 12 \lambda_2^2+4 \lambda_3^2+4 \lambda_3 \lambda_4+2 \lambda_4^2+2 \lambda_5^2+\dfrac{3}{4}\l(\dfrac{9}{25} g_1^4+\dfrac{6}{5} g_1^2g_2^2+3g_2^4\r) \\ & - 3 \lambda_2\l(\dfrac{3}{5} g_1^2+3g_2^2\r) +4 \lambda_2 T_\nu -4 T_{4\nu};\\
16 \pi^2 \dfrac{d \lambda_3}{d t} =& 2 (\lambda_1+\lambda_2)(3 \lambda_3+\lambda_4)+4 \lambda_3^2+2 \lambda_4^2+2 \lambda_5^2+\dfrac{3}{4}\l(\dfrac{9}{25} g_1^4-\dfrac{6}{5} g_1^2g_2^2+3g_2^4\r) \\ & - 3 \lambda_3\l(\dfrac{3}{5}g_1^2+3g_2^2\r) +2 \lambda_3 (T+T_\nu) -4 T_{\nu e};
\end{split}
\end{equation}
\begin{equation}
\begin{split}
16 \pi^2 \dfrac{d \lambda_4}{d t} =& 2 (\lambda_1+\lambda_2)\lambda_4+8 \lambda_3\lambda_4+4 \lambda_4^2+8 \lambda_5^2+\dfrac{9}{5} g_1^2g_2^2 \\ & - 3 \lambda_4\l(\dfrac{3}{5} g_1^2+3g_2^2\r) +2 \lambda_4 (T+T_\nu) +4 T_{\nu e};\\
16 \pi^2\dfrac{d \lambda_5}{d t} =& \lambda_5 \[ 2 (\lambda_1+\lambda_2)+8 \lambda_3+12 \lambda_4 - 3 \l(\dfrac{3}{5} g_1^2+3g_2^2\r) +2  (T+T_\nu) \];\\
\end{split}
\end{equation}
\begin{equation}
\begin{split}
\text{where}\hspace{1cm}T_{\nu} \equiv & \tr \[ \mathbf{h^\dagger h}\]; \hspace{.5cm}
T_{4\nu} \equiv  \tr \[ \mathbf{h^\dagger h h^\dagger h}\];\hspace{.5cm}
T_{\nu e} \equiv  \tr \[ \mathbf{h^\dagger h Y_e^\dagger Y_e}\].\\
\end{split}
\end{equation}

The RGEs for the mass parameters are given by Eq. (\ref{mass RGEs for IDM}). One notices from the set of RGEs that the evolution of Majorana Mass $(\mathbf{M})$, the new Yukawa coupling $(\mathbf{h})$ and the scalar quartic coupling $\lambda_5$ are proportional to the respective quantities themselves. The upshot of this setting is that these parameters remain small if they are small at any energy scale. This feature of the model becomes self-explanatory upon realization that if any of these parameters becomes zero, the neutrino becomes massless and global $U(1)$  symmetry conserving the lepton number is restored. \\

\section{Complete set of RGEs for scalar singlet dark matter model}
\label{appendix: RGEs for Scalar Extention}
While the RGEs for the mass parameters are given by Eq. (\ref{mass RGEs SM +real singlet scalar}), the RGEs for the quartic couplings are given by:
\begin{equation}
\label{Eq:RGEs for SM extension}
\begin{split}
16 \pi^2 \dfrac{d \lambda_1}{d t}=& \; 12 \lambda_1^2 - 3 \lambda_1 \l(\dfrac{3}{5}g_1^2+3g_2^2 \r) +\dfrac{3}{2} g_2^4+\dfrac{3}{4}\l(g_2^2+\dfrac{3}{5}g_1^2 \r)^2+4\lambda_1 T- 4 H+\dfrac{\lambda_3^2}{2};\\
16 \pi^2 \dfrac{d \lambda_2}{d t}=& \; 3\lambda_2^2+\dfrac{4}{3} \lambda_3^2;\\
16 \pi^2 \dfrac{d \lambda_3}{d t}=& \; 6 \lambda_3 (\lambda_1+\lambda_2).\\
\end{split}
\end{equation}
RGEs for the Yukawa couplings and the gauge couplings remain the same as SM.

\section{Complete set of RGEs for vector-like fermion model}
\label{appendix: RGEs for vector-like Fermion}
The RGEs for the mass parameters are given by Eq. (\ref{mass RGEs for vector-like Fermion model}). The RGEs for the gauge couplings are given by:
\begin{equation}
16 \pi^2 \dfrac{d g_i}{d t}=b_i g_i^3,
\end{equation}
where $b_i=\{-3,\dfrac{-19}{6},\dfrac{105}{10},\dfrac{259}{3}\}$.

The set of RGEs for all the Yukawa couplings is given by:
\begin{equation}
\begin{split}
16 \pi^2 \dfrac{d\mathbf{{Y}_u}}{d t}=&\; \mathbf{Y_u} \[  \dfrac{3}{2} \l( \mathbf{Y_u^\dagger Y_u} -\mathbf{Y_d^\dagger Y_d} \r)+\dfrac{1}{2}\mathbf{G_u^\dagger G_u}+ T- \dfrac{17}{20}g_1^2-\dfrac{9}{4}g_2^2-8g_3^2-3g_4^2\];\\
16 \pi^2 \dfrac{d\mathbf{{Y}_d}}{d t}=&\; \mathbf{Y_d} \[  \dfrac{3}{2} \l( \mathbf{Y_d^\dagger Y_d} - \mathbf{Y_u^\dagger Y_u} \r)+\dfrac{1}{2}\mathbf{G_d^\dagger G_d}+ T- \dfrac{1}{4}g_1^2-\dfrac{9}{4}g_2^2-8g_3^2-3g_4^2\];\\
16 \pi^2 \dfrac{d\mathbf{{Y}_e}}{d t}=&\; \mathbf{Y_e} \[  \dfrac{3}{2}  \mathbf{Y_e^\dagger Y_e} +\dfrac{1}{2}\mathbf{G_e^\dagger G_e}+ T- \dfrac{9}{4}g_1^2-\dfrac{9}{4}g_2^2-3g_4^2\];\\
16 \pi^2\dfrac{d\mathbf{{F}_u}}{d t}=& \; \mathbf{F_u} \[\mathbf{F_u^\dagger F_u}+T_F-\dfrac{8}{5}g_1^2-8g_3^2-15 g_4^2 \]+\dfrac{1}{2}\mathbf{G_u G_u^\dagger F_u};\\
16 \pi^2 \dfrac{d\mathbf{{F}_d}}{d t}=& \; \mathbf{F_d} \[\mathbf{F_d^\dagger F_d}+T_F-\dfrac{2}{5}g_1^2-8g_3^2-15 g_4^2 \]+\dfrac{1}{2}\mathbf{G_d G_d^\dagger F_d};\\
16 \pi^2 \dfrac{d\mathbf{{F}_e}}{d t}=& \; \mathbf{F_e} \[\mathbf{F_e^\dagger F_e}+T_F-\dfrac{18}{5}g_1^2-15 g_4^2 \]+\dfrac{1}{2}\mathbf{G_e G_e^\dagger F_e};\\
16 \pi^2 \dfrac{d\mathbf{{G}_u}}{d t}=& \; \mathbf{G_u} \[\mathbf{G_u^\dagger G_u}+\mathbf{Y_u^\dagger Y_u}+T_G-\dfrac{8}{5}g_1^2-8 g_3^2-6 g_4^2 \]+\dfrac{1}{2}\mathbf{F_u F_u^\dagger G_u};\\
16 \pi^2 \dfrac{d\mathbf{{G}_d}}{d t}=& \; \mathbf{G_d} \[\mathbf{G_d^\dagger G_d}+\mathbf{Y_d^\dagger Y_d}+T_G-\dfrac{2}{5}g_1^2-8 g_3^2-6 g_4^2 \]+\dfrac{1}{2}\mathbf{F_d F_d^\dagger G_d};\\
16 \pi^2\dfrac{d\mathbf{{G}_e}}{d t}=& \; \mathbf{G_e} \[\mathbf{G_e^\dagger G_e}+\mathbf{Y_e^\dagger Y_e}+T_G-\dfrac{18}{5}g_1^2-6 g_4^2 \]+\dfrac{1}{2}\mathbf{F_e F_e^\dagger G_e}.\\
\end{split}
\end{equation}
The RGEs for scalar quartic couplings are given by:
\begin{equation}
\begin{split}
16 \pi^2 \dfrac{d \lambda_1}{d t}=&\; 12 \lambda_1^2+2\lambda_4^2+2 \lambda_5^2-3 \lambda_1 \l( \dfrac{3}{5}g_1^2+3 g_2^2+4g_4^2 \r)+\l( \dfrac{27}{100} g_1^4+\dfrac{9}{4}g_2^4+\dfrac{9}{10}g_1^2g_2^2\r) \\& +12 g_4^2+6g_2^2g_4^2+\dfrac{18}{5}g_1^2g_4^2+4 \lambda_1 T-4 H;\\
16 \pi^2 \dfrac{d \lambda_2}{d t}=&\; 10 \lambda_2^2+4 \lambda_4^2+2 \lambda_6^2-12\lambda_2 g_4^2+12g_4^4+\lambda_2 T_F -4H_F;\\
16 \pi^2\dfrac{d \lambda_3}{d t}=&\; 10 \lambda_3^2+4 \lambda_5^2+2 \lambda_6^2-48\lambda_3 g_4^2+48 g_4^4+4 \lambda_3 T_G -4H_G;\\
16 \pi^2 \dfrac{d \lambda_4}{d t}=&\; 6 \lambda_1 \lambda_4+2 \lambda_2 \lambda_4+2\lambda_5 \lambda_6+4\lambda_4^2-\lambda_4 \l( \dfrac{9}{2}g_2^2+\dfrac{9}{10}g_1^2+12 g_4^2\r) + 12g_4^4+2 \lambda_4 (T+T_F);
\end{split}
\end{equation}
\begin{equation}
\begin{split}
16 \pi^2 \dfrac{d \lambda_5}{d t}=&\; 6 \lambda_1 \lambda_5+4 \lambda_3 \lambda_5+2\lambda_4 \lambda_6+4\lambda_5^2-\lambda_5 \l( \dfrac{9}{2}g_2^2+\dfrac{9}{10}g_1^2+30 g_4^2\r) + 48g_4^4+2 \lambda_5 (T+T_G)-4H_{YG};\\ 
16 \pi^2 \dfrac{d \lambda_6}{d t}=&\; 4 \lambda_2 \lambda_6+4 \lambda_3 \lambda_6+4\lambda_4 \lambda_5+4\lambda_6^2-30\lambda_6 g_4^2 + 48g_4^4+2 \lambda_6 (T_F+T_G)-4H_{FG};\\ 
\end{split}
\end{equation}
where
\begin{equation}
\begin{split}
T &=\tr \[\mathbf{Y_e^\dagger Y_e}  +3 \mathbf{Y_d^\dagger Y_d} +3 \mathbf{Y_u^\dagger Y_u} \] ;\\
T_F &=\tr \[\mathbf{F_e^\dagger F_e}  +3 \mathbf{F_d^\dagger F_d} +3 \mathbf{F_u^\dagger F_u} \]; \\
T_G &=\tr \[\mathbf{G_e^\dagger G_e}  +3 \mathbf{G_d^\dagger G_d} +3 \mathbf{G_u^\dagger G_u} \]; \\
H &= \tr \[\mathbf{Y_e^\dagger Y_e Y_e^\dagger Y_e}  +3 \mathbf{Y_d^\dagger Y_d Y_d^\dagger Y_d} +3 \mathbf{Y_u^\dagger Y_u Y_u^\dagger Y_u} \];\\
H_F &= \tr \[\mathbf{F_e^\dagger F_e F_e^\dagger F_e}  +3 \mathbf{F_d^\dagger F_d F_d^\dagger F_d} +3 \mathbf{F_u^\dagger F_u F_u^\dagger F_u} \];\\
H_G &= \tr \[\mathbf{G_e^\dagger G_e G_e^\dagger G_e}  +3 \mathbf{G_d^\dagger G_d G_d^\dagger G_d} +3 \mathbf{G_u^\dagger G_u G_u^\dagger G_u} \];\\
H_{YG} &= \tr \[\mathbf{Y_e^\dagger Y_e G_e^\dagger G_e}  +3 \mathbf{Y_d^\dagger Y_d G_d^\dagger G_d} +3 \mathbf{Y_u^\dagger Y_u G_u^\dagger G_u} \];\\
H_{FG} &= \tr \[\mathbf{F_e^\dagger F_e G_e^\dagger G_e}  +3 \mathbf{F_d^\dagger F_d G_d^\dagger G_d} +3 \mathbf{F_u^\dagger F_u G_u^\dagger G_u} \].
\end{split}
\end{equation} 
\end{appendices}

\end{document}